\documentclass{emulateapj}

\usepackage{graphicx}

\begin{document}

\title{GRB\,081007 and GRB\,090424: the surrounding medium, outflows and supernovae}

\def\INAFBrera{1}
\def\PMOCAS{2}
\def\INAFNapoli{3}
\def\ICRAN{4}
\def\IAC{5}
\def\ULL{6}
\def\Dark{7}
\def\SNS{8}
\def\INAFIASFBO{9}
\def\INAFMilano{10}
\def\LJMU{11}
\def\CACUI{12}
\def\IAACSIC{13}
\def\ULjubljana{14}
\def\UPVEHU{15}
\def\Ikerbasque{16}
\def\UFerrara{17}
\def\UNC{18}
\def\ESO{19}
\def\MPI{20}
\def\INAFPadova{21}
\def\INAFRoma{22}
\def\Leicester{23}
\def\LANL{24}
\def\Yale{25}

\author{
Zhi-Ping Jin\altaffilmark{\INAFBrera,\PMOCAS},
Stefano Covino\altaffilmark{\INAFBrera},
Massimo Della Valle\altaffilmark{\INAFNapoli,\ICRAN},
Patrizia Ferrero\altaffilmark{\IAC,\ULL}, 
Dino Fugazza\altaffilmark{\INAFBrera}, 
Daniele Malesani\altaffilmark{\Dark},
Andrea Melandri\altaffilmark{\INAFBrera},
Elena Pian\altaffilmark{\SNS,\INAFIASFBO},
Ruben Salvaterra\altaffilmark{\INAFMilano}, 
David Bersier\altaffilmark{\LJMU}, 
Sergio Campana\altaffilmark{\INAFBrera}, 
Zach Cano\altaffilmark{\CACUI}, 
Alberto J. Castro-Tirado\altaffilmark{\IAACSIC}, 
Paolo D'Avanzo\altaffilmark{\INAFBrera},
Johan P. U. Fynbo\altaffilmark{\Dark},
Andreja Gomboc\altaffilmark{\ULjubljana},
Javier Gorosabel\altaffilmark{\IAACSIC,\UPVEHU,\Ikerbasque}, 
Cristiano Guidorzi\altaffilmark{\UFerrara},  
Joshua B. Haislip\altaffilmark{\UNC}, 
Jens Hjorth\altaffilmark{\Dark},  
Shiho Kobayashi\altaffilmark{\LJMU}, 
Aaron P. LaCluyze\altaffilmark{\UNC},
Gianni Marconi\altaffilmark{\ESO}, 
Paolo A. Mazzali\altaffilmark{\LJMU,\MPI,\INAFPadova}, 
Carole G. Mundell\altaffilmark{\LJMU}, 
Silvia Piranomonte\altaffilmark{\INAFRoma}, 
Daniel E. Reichart\altaffilmark{\UNC}, 
Rub\'en S\'anchez-Ram\'{i}rez\altaffilmark{\IAACSIC}, 
Robert J. Smith\altaffilmark{\LJMU}, 
Iain A. Steele\altaffilmark{\LJMU}, 
Gianpiero Tagliaferri\altaffilmark{\INAFBrera},
Nial R. Tanvir\altaffilmark{\Leicester}, 
Stefano Valenti\altaffilmark{\INAFPadova}, 
Susanna D. Vergani\altaffilmark{\INAFBrera},
Thomas Vestrand\altaffilmark{\LANL}, 
Emma S. Walker\altaffilmark{\SNS,\Yale} and
Przemek Wo{\'z}niak\altaffilmark{\LANL}
}

\altaffiltext{1}{
INAF-Osservatorio Astronomico di Brera, via Emilio Bianchi 46, I-23807 Merate (LC), Italy
\label{INAF-Brera}}

\altaffiltext{2}{
Key Laboratory of Dark Matter and Space Astronomy, Purple Mountain Observatory, Chinese Academy of Sciences, Nanjing 210008, China \email{jin@pmo.ac.cn}
\label{PMO-CAS}}

\altaffiltext{3}{
INAF-Osservatorio Astronomico di Capodimonte, Salita Moiariello 16, I-80131 Napoli, Italy
\label{INAF-Napoli}}

\altaffiltext{4}{
International Center for Relativistic Astrophysics Network, Piazza della Repubblica 10, I-65122 Pescara, Italy
\label{ICRAN}}

\altaffiltext{5}{
Instituto de Astrof\'isica de Canarias (IAC), E-38200 La Laguna, Tenerife, Spain
\label{IAC}}

\altaffiltext{6}{
Departamento de Astrof\'isica, Universidad de La Laguna (ULL), E-38205 La Laguna, Tenerife, Spain
\label{ULL}}

\altaffiltext{7}{
Dark Cosmology Centre, Niels Bohr Institute, University of Copenhagen, Juliane Maries Vej 30, DK-2100 Copenhagen \O, Denmark
\label{Dark}}

\altaffiltext{8}{
Scuola Normale Superiore di Pisa, Piazza dei Cavalieri 7, I-56126 Pisa, Italy
\label{SNS}}

\altaffiltext{9}{
INAF-Istituto di Astrofisica Spaziale e Fisica Cosmica, via P. Gobetti 101, I-40129 Bologna, Italy
\label{INAF-IASFBO}}

\altaffiltext{10}{
INAF-IASF Milano, via E. Bassini 15, I-20133 Milano, Italy
\label{INAF-Milano}}

\altaffiltext{11}{
Astrophysics Research Institute, Liverpool John Moores University, Liverpool Science Park, 146 Brownlow Hill, Liverpool L3 5RF, UK
\label{LJMU}}

\altaffiltext{12}{
Centre for Astrophysics and Cosmology, Science Institute, University of Iceland, Reykjavik, Iceland
\label{UACUI}}

\altaffiltext{13}{
Instituto de Astrof\' isica de Andaluc\' ia (IAA-CSIC), Glorieta de la Astronom\' ia s/n, E-18008, Granada, Spain
\label{IAA-CSIC}}

\altaffiltext{14}{
Mathematics \& Physics, University of Ljubljana, Jadranska ulica 19, 1000 Ljubljana, Slovenia
\label{U.Ljubljana}}

\altaffiltext{15}{
Unidad Asociada Grupo Ciencia Planetarias UPV/EHU-IAA/CSIC, Departamento de F\'{\i}sica Aplicada I, E.T.S. Ingenier\'{\i}a, Universidad del Pa\'{\i}s Vasco UPV/EHU, Alameda de Urquijo s/n, E-48013 Bilbao, Spain
\label{UPV/EHU-IAA/CSIC}}

\altaffiltext{16}{
Ikerbasque, Basque Foundation for Science, Alameda de Urquijo 36-5, E-48008 Bilbao, Spain
\label{Ikerbasque}}

\altaffiltext{17}{
Department of Physics, University of Ferrara, via Saragat 1, I-44122 Ferrara, Italy
\label{U.Ferrara}}

\altaffiltext{18}{
Department of Physics and Astronomy, University of North Carolina, Chapel Hill NC 27599, USA
\label{UNC}}

\altaffiltext{19}{
European Southern Observatory, Casilla 19001, Santiago, Chile
\label{ESO}}

\altaffiltext{20}{
Max-Planck-Institut f\"ur Astrophysik, Karl-Schwarzschildstr. 1, D-85748 Garching, Germany
\label{MPI}}

\altaffiltext{21}{
INAF-Osservatorio Astronomico di Padova, Vicolo dell'Osservatorio 5, I-35122 Padova, Italy
\label{INAF-Padova}}

\altaffiltext{22}{
INAF-Osservatorio Astronomico di Roma, via Frascati 33, I-00040 Monte Porzio Catone, Roma, Italy
\label{INAF-Roma}}

\altaffiltext{23}{
Department of Physics and Astronomy, University of Leicester, Leicester LE1 7RH, UK
\label{Leicester}}

\altaffiltext{24}{
Space and Remote Sensing, Los Alamos National Laboratory, MS-B244, Los Alamos, NM 87545, USA
\label{LANL}}

\altaffiltext{25}{
Yale University, Department of Physics, 217 Prospect Street, New Haven, CT 06511, USA
\label{Yale}}

\begin{abstract}
We discuss the results of the analysis of multi-wavelength data for the afterglows of GRB 081007 and GRB 090424, 
two bursts detected by \textit{Swift}. One of them, GRB 081007, also shows a spectroscopically confirmed supernova, 
SN 2008hw, which resembles SN 1998bw in its absorption features, while the maximum magnitude may be fainter, up to 0.7 mag,  than observed in SN 1998bw. 
Bright optical flashes have been detected in both events, which allows us to derive solid
constraints on the circumburst-matter density profile. 
This is particularly interesting in the case of GRB\,081007, 
whose afterglow is found to be propagating into a constant-density medium, 
yielding yet another example of a gamma-ray burst (GRB) clearly associated with a massive star progenitor 
which did not sculpt the surroundings with its stellar wind. 
There is no supernova component detected in the afterglow of GRB\,090424, likely because of
the brightness of the host galaxy, comparable to the Milky Way. We show
that the afterglow data are consistent with the presence of both forward- and
reverse-shock emission powered by relativistic outflows expanding into the
interstellar medium. The absence of optical peaks due to the forward shock
strongly suggests that the reverse-shock regions should be mildly magnetized.
The initial Lorentz factor of outflow of GRB\,081007 is estimated to be $\Gamma
\sim 200$, while for GRB\,090424 a lower limit of $\Gamma>170$ is derived. We also
discuss the prompt emission of GRB\,081007, which consists of just a
single pulse. We argue that neither the external forward-shock
model nor the shock-breakout model can account for the prompt emission data and suggest that
the single-pulse-like prompt emission may be due to magnetic energy dissipation
of a Poynting-flux-dominated outflow or to a dissipative photosphere. 
\end{abstract}



\section{Introduction}

Thanks to the rapid localization of gamma-ray bursts (GRBs) by the
\textit{Swift} satellite \citep{Gehrels04}, the response of ground-based
observations of GRB afterglows has been greatly enhanced and observations of GRB
afterglows have become routinely possible within minutes after the explosion.
Very early afterglow data are required to estimate the initial bulk Lorentz
factor of the ejecta \citep{Molinari07}, probe the physical composition of the
outflow \citep{Fan02,Zhang03}, and constrain the density profile of the medium
surrounding the progenitor \citep{Jin07,Schulze11}. Late afterglow observations provide us
with clues on the medium density profile, which, in turn, may provide hints on the
nature of the progenitor star. 

The idea that supernova (SN) explosions may also produce energetic gamma-ray emission 
by some mechanism goes back to \citet{Colgate68}, 
but the first piece of evidence supporting such a connection was not found until 1998
\citep{Galama98}. The connection was finally established in 2003 when SN\,2003dh and SN\,2003lw 
were unambiguously detected spectroscopically following the nearby GRB\,030329 and GRB\,031203, 
respectively \citep{Hjorth03,Stanek03,Malesani04}. 
So far, spectral SN features have been found for only about a dozen GRBs \citep[for  recent
reviews see][]{Hjorth12,DellaValle11}. Current data suggest that less than $3\%$
of Type Ib/c SNe are able to produce GRBs following the core collapse of their
progenitor star \citep{Guetta07,Soderberg10,Ghirlanda13}.

In this work we present and discuss data of GRB\,081007 and GRB\,090424.  Because of
their occurrence at relatively low redshifts, $0.53$ and $0.54$, respectively,
their follow-up in the optical and near-infrared (NIR) bands was particularly 
effective \citep[see cases of GRBs at similar redshifts reported by][]{DellaValle06,Cano11,Berger11,Filgas11,Troja12,Sparre11}. 
We were able to detect a SN component in the late afterglow of GRB\,081007. 

\begin{figure}
\includegraphics[width=0.5\textwidth]{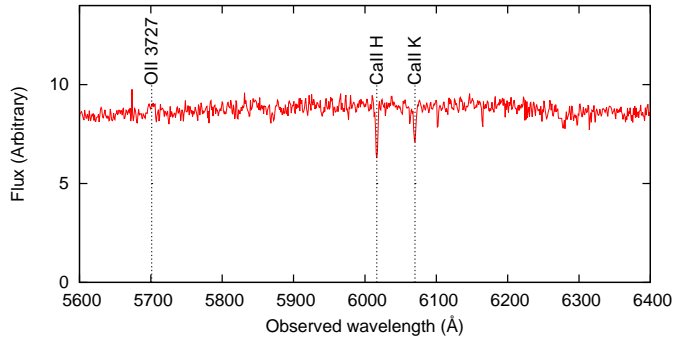}
\caption{Gemini-South GMOS spectrum of GRB 081007 afterglow.}
\label{GRB081007s}
\end{figure}

GRB\,081007 triggered the Burst Alert Telescope (BAT) onboard \textit{Swift} on 2008 October 7 at
05:23:52 UT. It was a long GRB with a peak $\sim9$~s duration in the 15-350 keV
gamma-ray band \citep{Markwardt08}. This burst also triggered the Gamma-ray
Burst Monitor, onboard \textit{Fermi}, in the 25-900 keV gamma-ray band. 
The prompt emission consisted of a single pulse with an estimated duration
$T_{90}$ of about 12~s \citep{Bissaldi08}. The \textit{Swift} satellite
immediately slewed to the field and the X-Ray Telescope (XRT) and the
Ultraviolet/Optical Telescope (UVOT) started observations at 99 and 108~s
after the trigger, leading to a detection of the X-ray and optical afterglows
\citep{Baumgartner08}. Gemini-South took two 900~s spectra, starting 73 minutes
after the burst. Two absorption lines at 6016.7 and 6070.3 \AA{} and an emission
line at 5700.9 \AA{} were identified as \ion{Ca}{2} H, K and [\ion{O}{2}] 3727
\AA{} lines respectively at $z = 0.5295\pm0.0001$ \citep[][see also figure \ref{GRB081007s}]{Berger08}.

\begin{figure}
\includegraphics[width=0.5\textwidth]{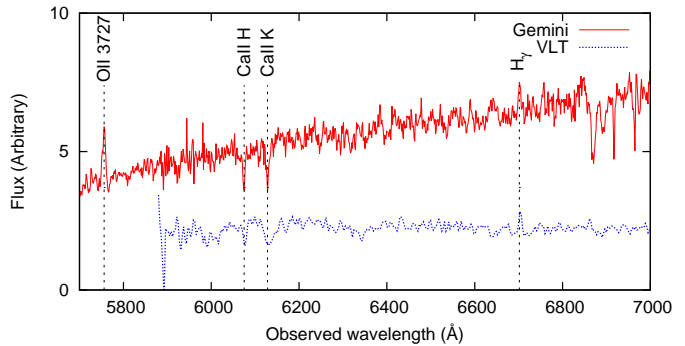}
\caption{Gemini-South GMOS spectrum of GRB 090424 afterglow and the VLT FORS spectrum of its host galaxy.}
\label{GRB090424s}
\end{figure}

GRB\,090424 triggered BAT on 2009 April 24 at 14:12:09 UT. It was a multi-pulse
long GRB with total duration of about 60~s in the 15-350 keV gamma-ray band
\citep{Cannizzo09}. XRT and UVOT started observations 85 and 91~s after the
trigger and detected the X-ray and optical afterglow \citep{Cannizzo09}.
Gemini-South took two 1200~s spectra, starting 11.7~hr after the burst, 
and found that GRB\,090424 is at $z=0.544$, similar to GRB\,081007 \citep[][see also figure \ref{GRB090424s}]{Chornock09}. 

The main properties of the two GRBs are
summarized in Table \ref{comparison}.
This work is structured as follows: in Section 2 we present the data. The
discussion and interpretation of the observations are reported in Section 3. We
summarize our results in Section 4.

\begin{table}
\centering
\caption{Properties of GRB\,081007 and GRB\,090424}
\begin{tabular}{l l l}
\hline
Name                 & GRB\,081007        & GRB\,090424            \\
\hline
$T_{90}$ (s)         & 8.0 (1)            & $48 \pm3$ (2)          \\
$z$                  & $0.5295\pm0.0001$  & 0.544                  \\
$E_{\rm peak}$ (keV) & $61\pm15$ (1)      & $236^{+127}_{-49}$ (3) \\
$E_{\gamma,\rm iso}$ (erg) & $1.5_{-0.3}^{+0.4}\times10^{51}$ (1) & $4.3^{+2.4}_{-1.4}\times10^{52}$ (3) \\
$\Gamma$             & $\sim 200$         & $> 170$                \\
$R_{\rm B}$          & $\sim10{R_{\rm e}^{-0.75}}$ & $>10{R_{\rm e}^{-0.72}}$\\
SN                   & SN\,2008hw         & No detection           \\
$R_{\rm galaxy}$     & $>24.63$              & $22.01\pm0.12$         \\
\hline
\end{tabular}
\tablerefs{(1) \citet{deUgartePostigo11}, (2) \citet{Sakamoto09}, (3) \citet{Sakamoto11}.}
\label{comparison}
\end{table}

\section{Observations and Results}

\subsection{The GRB\,081007 and GRB\,090424 afterglows}

Many robotic telescopes reacted to the trigger from GRB\,081007. RAPTOR started
observations detecting the optical counterpart after about 24\,s in the $R$ band
\citep{Wren08}. Four of the 40 cm PROMPT telescopes at CTIO in Chile began
observing after 41\,s \citep{West08}. The 60 cm Rapid Eye Mount (REM) 
on La Silla, Chile, started multi-band observations of GRB\,081007 after 46\,s 
in the $R$, $H$ and $K_{\rm s}$ bands \citep{Covino08}. The 2 m Faulkes
Telescope North (FTN; Haleakala, Hawaii) started to observe GRB\,081007 about
17\,minutes after the \textit{Swift} trigger 
in the $B$, $V$, $R$, and $I$ bands \citep{Smith08}. The 2 m Faulkes Telescope
South (FTS; Siding Spring, Australia) observed the field of GRB\,081007 from
$\sim1.1$ to $\sim4.2$ days after the burst in the $R$ and $I$ bands. 
Between 2008 October 24 and 2009 January 3, FORS2 at the Very Large Telescope (VLT)
was used to search for the SN associated with GRB\,081007. 

The 2 m Liverpool Telescope (LT; La Palma, Canary Islands, Spain) began
observing the field of GRB\,090424 at 21:29:47 UT \citep{Guidorzi09}, $\sim 7$
hr after the burst. The optical counterpart was clearly detected in the $r$ and
$i$ filters. Later observations were made by the FTN at 17.66 and 333.89 days
after the trigger in the $R$ filter only. Between 2009 May 1 and July 5, FORS2
at the VLT was also used to observe the GRB\,090424 field for nine runs. 
Three years later, this field was observed again between 2012 May 1 to 3 
with the 3.5m telescope at the Calar Alto Observatory. 

Our dataset also includes \textit{Swift}/UVOT data\footnote{The
\textit{Swift}/UVOT data are provided by the High Energy Astrophysics Science
Archive Research Center (HEASARC).} 
and we have retrieved and analyzed public Gemini archival data
\footnote{Gemini data are obtained at the Gemini Observatory, which is operated
    by the Association of Universities for Research in Astronomy.} 
of GRB\,081007 and GRB\,090424, confirming results reported by
\citet{Berger08} and \citet{Chornock09}.

In this paper we analyzed all available photometric and spectroscopic data,
following standard procedures: bias or dark removal, flat-field correction,
astrometry for imaging frames and wavelength calibration for spectroscopy. 
Aperture photometry was calibrated by means of secondary standard stars in the field from
the APASS catalog\footnote{http://www.aavso.org/apass} 
or the Sloan Digital Sky Survey (SDSS) catalog\footnote{http://www.sdss.org} in the optical 
and the Two Micron All Sky Survey (2MASS) catalog\footnote{http://www.ipac.caltech.edu/2mass} in the near-infrared
(NIR). R and I band observations were calibrated by means of  $r'$ and $i'$  secondary standard stars.
Late-time photometric observations, since all secondary standard stars in the field were heavily saturated,
were calibrated by means of standard star fields observed under photometric conditions.
Spectroscopic observations were also calibrated by observations of suitable spectro-photometric
standard stars. 

We also collected data available from these GCN circulars: 
8339 \citep{Cobb08} for GRB\,081007;
9224 \citep{Yuan09}, 
9236 \citep{Gorosabel09}, 
9239 \citep{Oksanen09}, 
9240 \citep{Urata09}, 
9245 \citep{Olivares09}, 
9246 \citep{Nissinen09}, 
9248 \citep{Im09a}, 
9253 \citep{Im09b}, 
9278 \citep{Roy09}, 
9305 \citep{Mao09}, 
9313 \citep{Cobb09},
9320 \citep{Rumyantsev09} for GRB\,090424.

The photometric data are shown in Figures \ref{GRB081007light curves} and
\ref{GRB090424light curves} and they are reported in the Appendix of the online
journal (Tables \ref{081007data} - \ref{090424data}).
The (small) Galactic extinction, $E(B-V)$,
of 0.016 and 0.025 mag \citep{Schlegel98} for GRB\,081007 and GRB\,090424,
respectively, was taken properly into account in the analysis. 

\begin{figure*}
\begin{center}
  \includegraphics[width=0.8\textwidth]{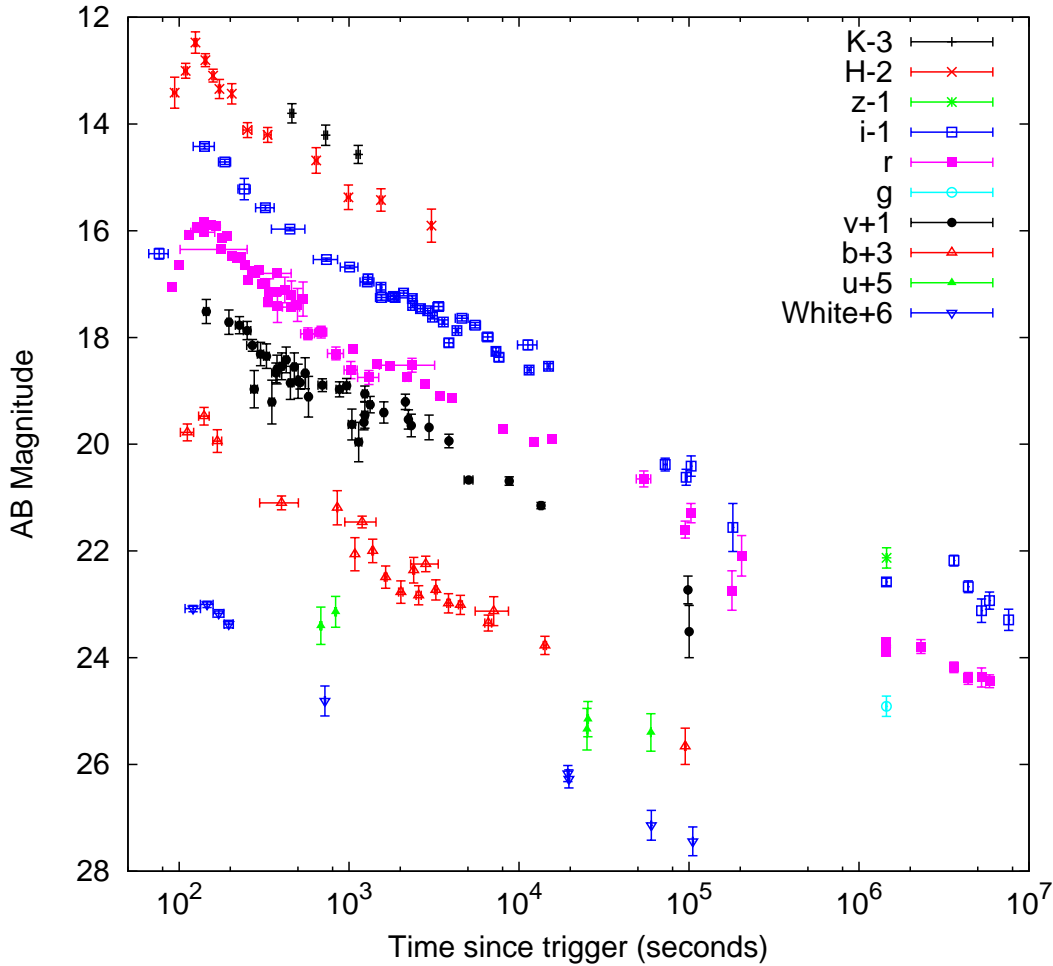}
\end{center}
  \caption{Optical/NIR observations of the GRB\,081007 afterglow. 
Similar filters (e.g., Johnson R and SDSS r) are plotted with the same symbols for clarity in the figure. 
The data can be found in Tables \ref{081007data} and 4 in the online material.}
  \label{GRB081007light curves}
\end{figure*}

\begin{figure*}
\begin{center}
  \includegraphics[width=0.8\textwidth]{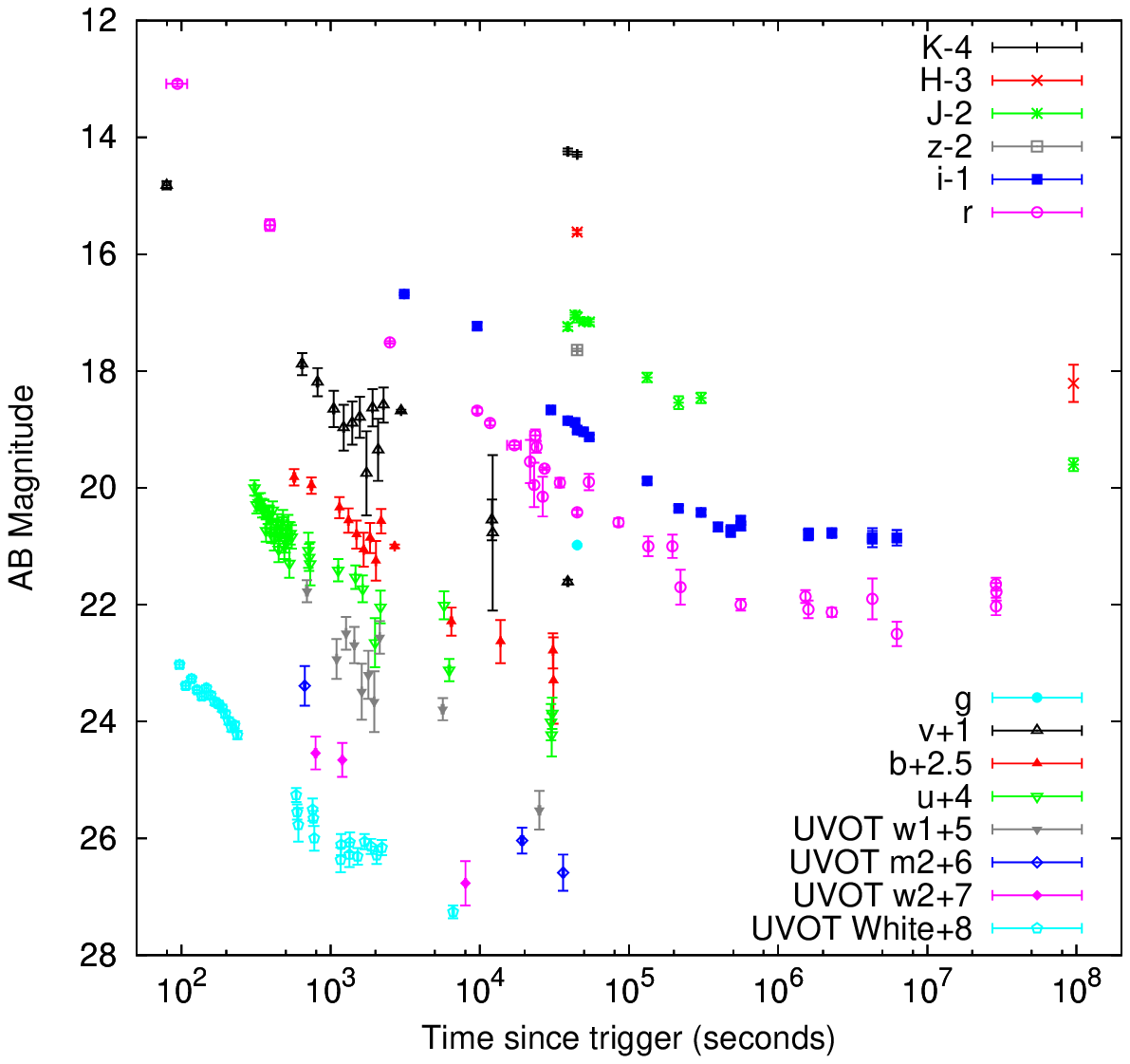}
\end{center}
  \caption{Optical/NIR observations of the GRB\,090424 afterglow. 
Similar filters (e.g., Johnson R and SDSS r) are plotted with same the symbols for clarity in the figure. 
The data can be found in Table \ref{090424data} in the online material.}
  \label{GRB090424light curves}
\end{figure*}

\subsection{Discovery of a supernova accompanying GRB081007}

\begin{figure}
  \includegraphics[width=0.5\textwidth]{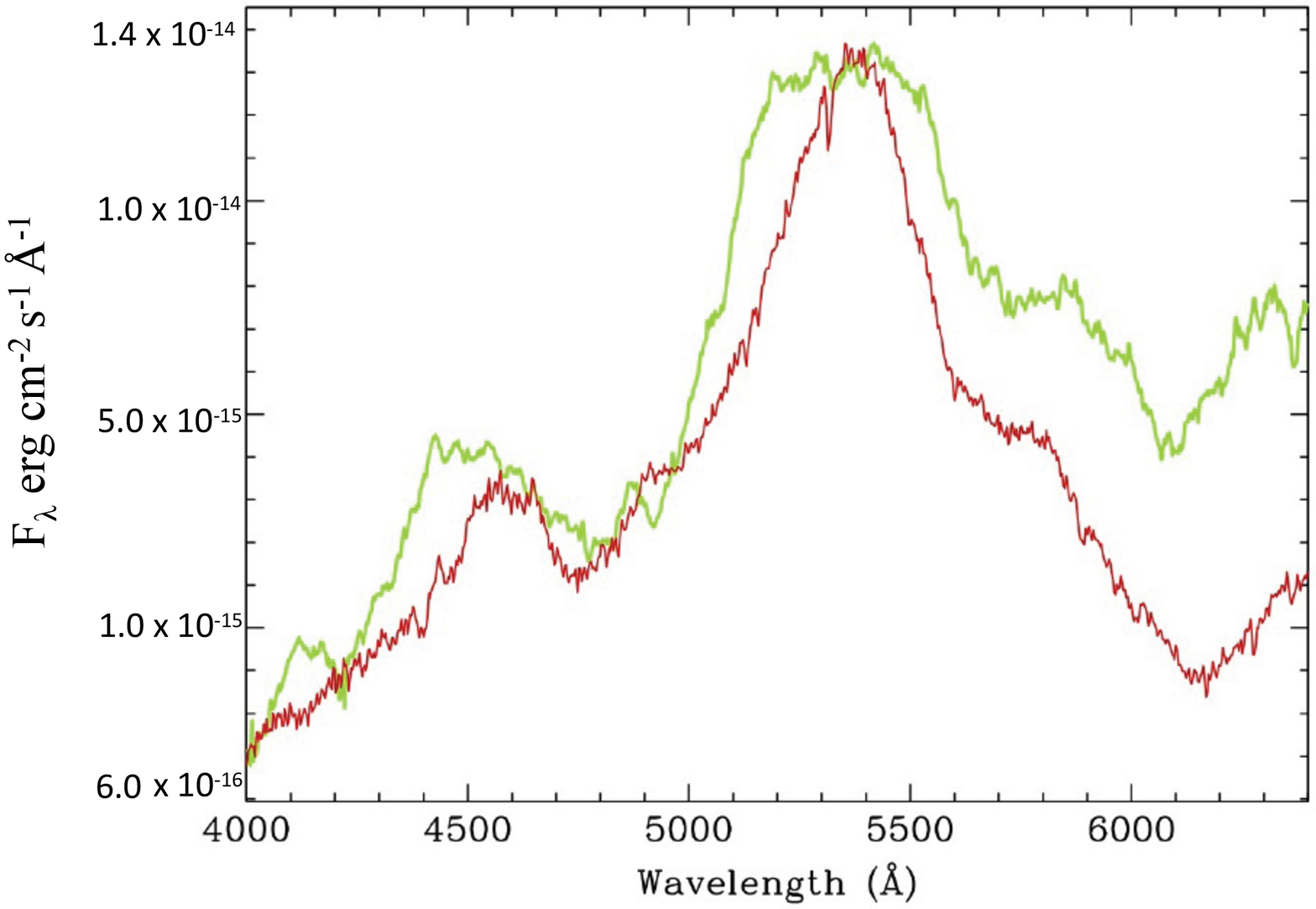}
  \caption{VLT-FORS2 spectrum of the optical counterpart of GRB 081007 taken on Nov 2, 2008, cleaned from telluric absorptions and 
  sky lines and smoothed with a boxcar of 30 \AA{} (green trace), compared with the spectrum of SN 1998bw close to maximum light 
  \citep[red trace, from][]{Patat01}. The similarity suggests the presence of a supernova underlying GRB081007, dubbed SN 2008hw. 
The spectrum of SN 2008hw was rescaled to best match the features of SN1998bw. }
  \label{sn2008hw}
\end{figure}

\begin{figure*}
\begin{center}
  \includegraphics[width=0.8\textwidth]{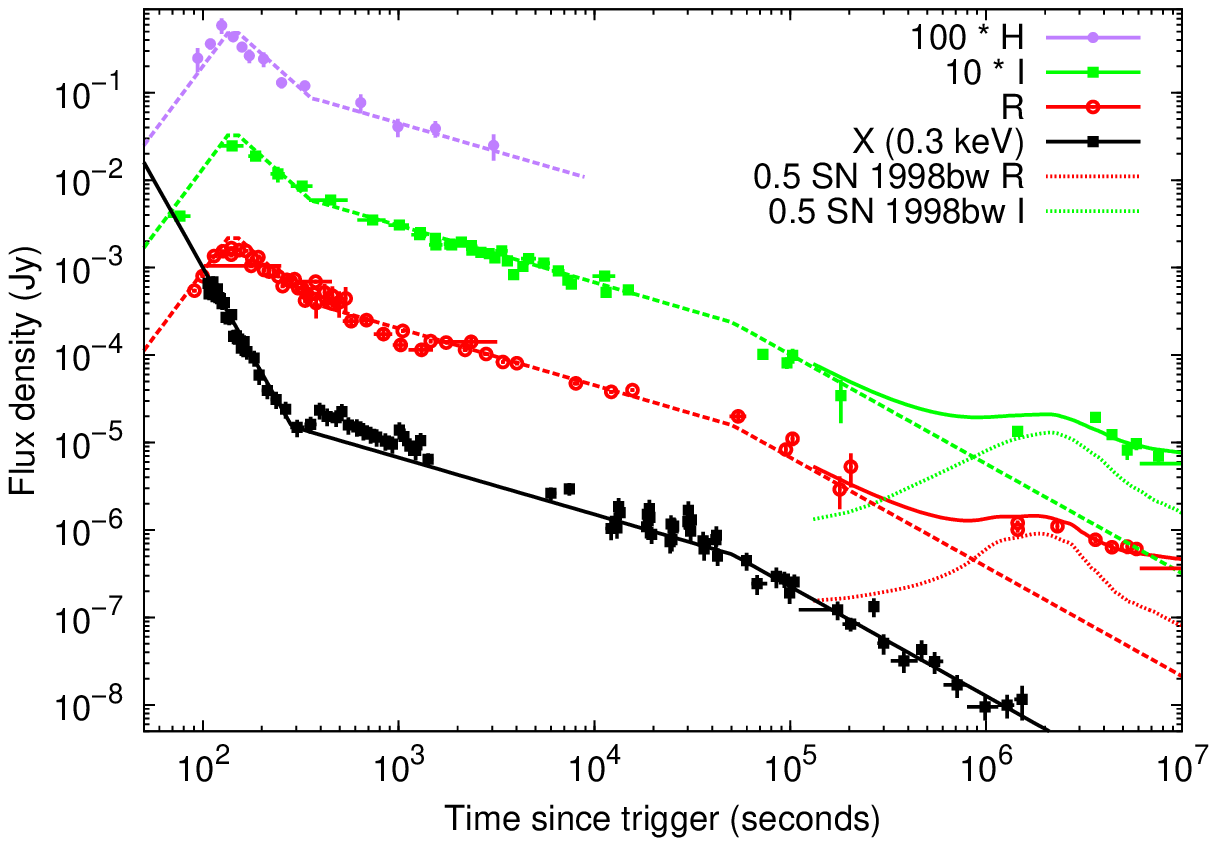}
\end{center}
\caption{Fits to the GRB\,081007 afterglow observations. 
The broken power-law lines with slope indices 3, $-2$, $-0.65$, and $-1.25$ for the optical/NIR
(long-dashed lines), and $-4$, $-0.65$, and $-1.25$ for the X-rays (black solid line) represent the afterglow light curves. 
The $R$ and $I$-band late observations can be interpreted as the sum (red and green solid lines respectively) of the power-law
afterglow, a SN template and a host-galaxy component, here the SN is 0.5 times as bright as SN\,1998bw,
reported at $z = 0.5295$, the host galaxy is $m_R = 25.0$ and $m_I = 24.5$ (short lines on the right). The SN template is also plotted
for comparison (dotted line). The \textit{Swift}/XRT X-ray light curve has been
retrieved from the UK \textit{Swift} Science Data Centre \citep{Evans09}.}
\label{081007fit}
\end{figure*}

A 2~hr spectrum of GRB\,081007 was obtained with VLT equipped with FORS2 and the 300I grism on 2008 November 2
\citep{DellaValle08}, about 26 days after the GRB trigger, and reduced following standard methods.   After subtracting a
starburst galaxy template \citep[Sb1 template from][]{Kinney96}, three broad
bumps at about 4600, 5400, and 6400 \AA{} emerge. These features are very
similar to those exhibited by SN\,1998bw \citep{Patat01} around maximum (see
Figure \ref{sn2008hw}), but the luminosity of the SN at maximum light is significantly lower than that of
SN\,1998bw at the same time, only about half as large as that of SN 1998bw (see Figure \ref{081007fit}). This SN was designated
SN\,2008hw \citep{DellaValle08}. Some similarities with other two broad-lined
type Ic SNe, namely SN\,1997ef \citep{Mazzali00} and SN\,2004aw
\citep{Taubenberger06} were also identified.

\subsection{Host galaxy of GRB\,090424}

\begin{figure*}
\begin{center}
  \includegraphics[width=0.8\textwidth]{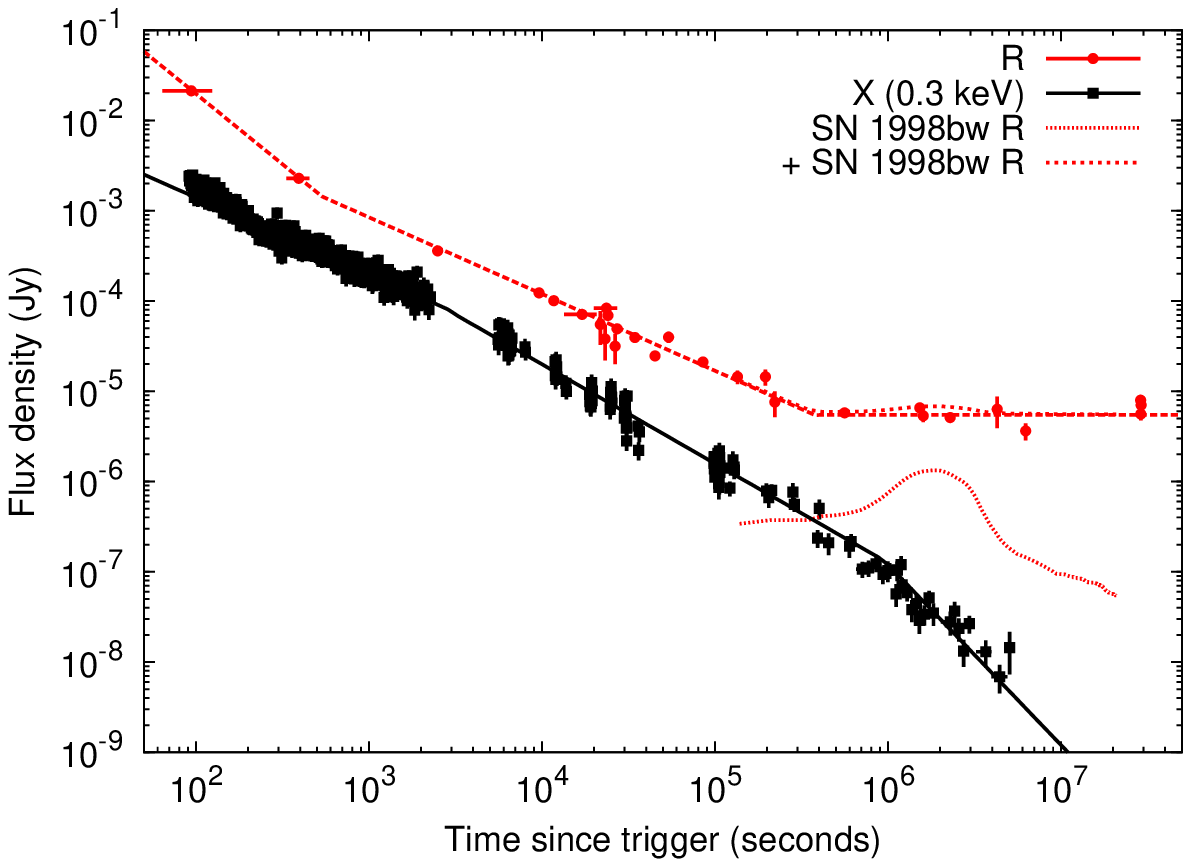}
\end{center}
\caption{Simple broken power-law fits to the observations of the GRB\,090424
afterglow. The slope indices are $-1.5$, $-0.85$ and 0 for the $R$ band (long
dashed line), $-0.85$, $-1.1$, and $-2$ for the X-rays (solid line). A smoothed
light curve of SN\,1998bw, shifted to $z = 0.544$, is plotted for comparison
(dotted line). An underlying SN like SN\,1998bw would contribute minimally to
the total flux (short dashed line), below the uncertainty of the luminosity
of the host galaxy.  The X-ray light curve has been derived using the
\textit{Swift}/XRT data from the UK \textit{Swift} Science Data Centre
\citep{Evans09}.} \label{090424fit}
\end{figure*}

The optical counterpart of GRB\,090424 shows no apparent variation from 18
days onwards, according to our FTN and VLT observations, no significant variation is found in \citet{Kann10} either. 
This means the afterglow
had faded below the host-galaxy brightness before this epoch. The VLT observed $r$
and $i$ band magnitudes of the host are $r = 22.07\pm 0.12$ and $i = 21.82 \pm
0.12$, corresponding to a luminosity four times brighter than SN\,1998bw at maximum
light. An underlying SN akin to SN\,1998bw would only have produced little additional 
brightening, at a level below the uncertainty in the galaxy luminosity. 
Additionally to the bright host galaxy, the afterglow suffers from significant host-frame extinction 
according to \citet{Kann10,Schady12} and \citet{Covino13}, which will further dim any SN component. 
This fact may explain the lack of detection
of the SN component in the afterglow of GRB\,090424 (see Figure
\ref{090424fit}).

\begin{figure*}
  \includegraphics[width=1.0\textwidth]{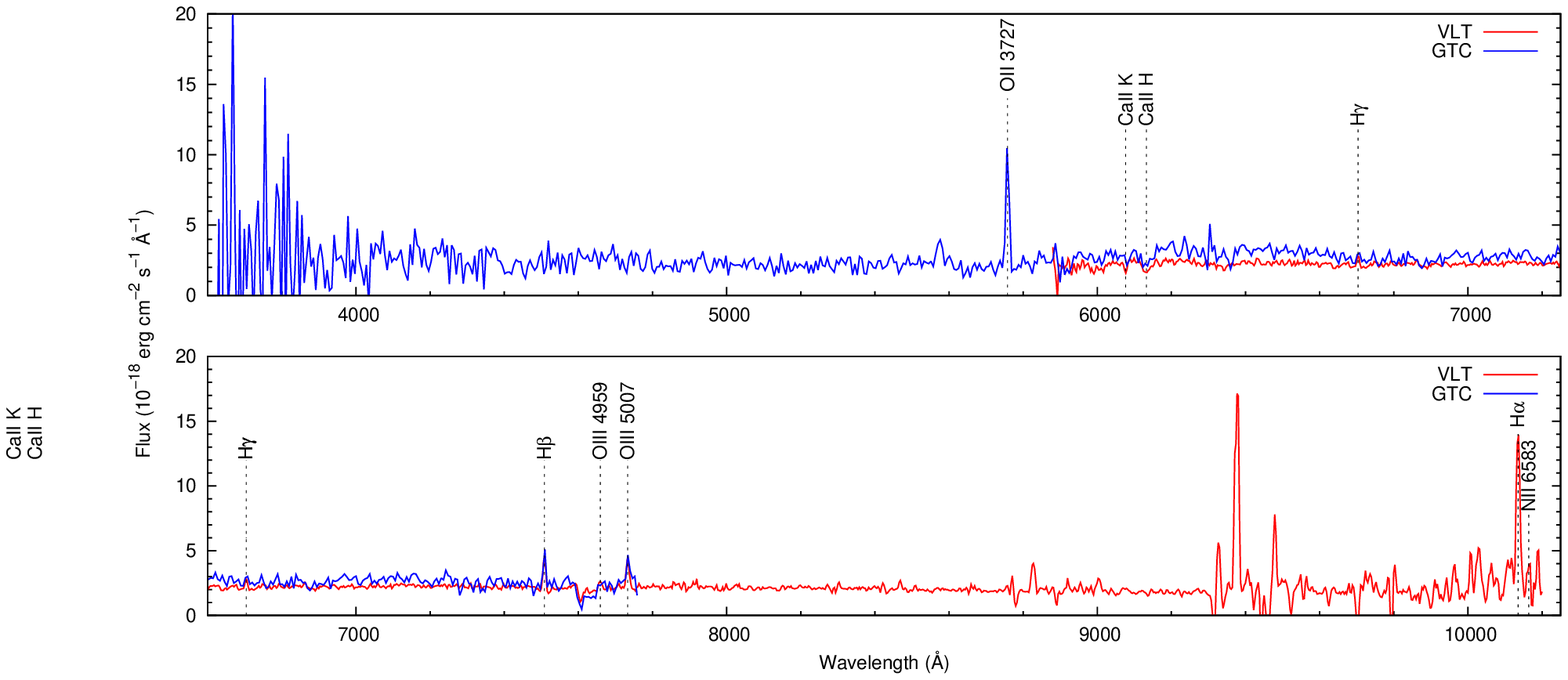}
  \caption{The VLT/FORS2 and the GTC spectra of the GRB\,090424 host galaxy in the observer frame.}
  \label{GRB090424host}
\end{figure*}

The GRB\,090424 host-galaxy spectrum was obtained with VLT-FORS2 and grism 300I on 2009 May 22, 
and with the Gran Telescopio Canarias (GTC) on 2013 April 8, 
about a month and 4 years after the high energy event respectively. 
These spectra were reduced following standard methods. The signal-to-noise ratio (S/N) of our VLT-FORS2 spectrum is $\sim10-20$ in 
the blue part and goes down to ~5-10 in the red part, the strong residual around 
9400\AA{} is due to telluric water vapor. Two absorption lines at 6077 and 6133
\AA{} and emission lines at 6704, 7508, 7733, and 10135 \AA{} are clearly
detected, which are identified as \ion{Ca}{2} K and H, H$\gamma$, H$\beta$,
[\ion{O}{3}] and H$\alpha$ at $z = 0.544$ (see Figure \ref{GRB090424host}). 
The S/N of our GTC spectrum is $\sim5-10$, it extends the VLT-FORS2 spectrum to the blue-ward, 
and an extra strong emission line at 5758 \AA{} is detected, it is identified as [\ion{O}{2}] at $z = 0.544$ (see Figure \ref{GRB090424host}). 
We derived the fluxes for these lines first by removing the background with 
a polynomial fit and then by modeling the lines with Gaussians. The resolution 
of our spectra does not justify a more sophisticated modeling. The results are 
summarized in Table \ref{hostlines}.

\begin{table}
\caption{Spectral lines of the GRB\,090424 host galaxy}
\begin{tabular}{lllllll}
\hline
Line          &$\lambda_{\rm rest}$ &$\lambda_{\rm obs}$ &FWHM           &$F$             \\ 
              &\AA                  &\AA                 &\AA            &$10^{-18}$ cgs  \\
\hline
{[}\ion{O}{2}{]} &3727.09              &5758.10             &$8.87\pm1.16$  &$82.25\pm13.82$ \\
\ion{Ca}{2} K &3934.78              &6076.66             &$5.83\pm1.84$  &$-3.73\pm1.59$  \\
\ion{Ca}{2} H &3969.59              &6132.57             &$16.81\pm3.02$ &$-11.01\pm2.62$ \\
H$\gamma$     &4341.68              &6703.95             &$7.59\pm0.44$  &$7.14\pm0.55$   \\
H$\beta$      &4862.68              &7508.30             &$7.94\pm0.63$  &$23.00\pm2.41$  \\
{[}\ion{O}{3}{]} &4960.30           &7659.25             &$9.94\pm1.27$  &$7.06\pm1.20$   \\
{[}\ion{O}{3}{]} &5008.24           &7733.10             &$9.90\pm0.43$  &$25.13\pm1.45$  \\
H$\alpha$     &6564.61              &10135.45            &$11.35\pm0.48$ &$151.64\pm8.45$ \\
\ion{N}{2}    &6585.27              &10164.24            &$8.33\pm4.02$  &$19.08\pm12.15$ \\
\hline
\end{tabular}
\tablecomments{The flux $F$ is as observed, but the Galactic extinction has been taken into account for the analysis.}
 \label{hostlines}
\end{table}

\section{Interpretation and implication of the data}

\subsection{Interpreting the GRB\,081007 afterglow}
\label{sec:081007aft}

The most interesting feature of the early REM light curve of GRB\,081007 is a
bright peak in the $H$ band data (see Figure \ref{081007fit}). The light curve
first rises as $t^{+3.0}$ and peaks at approximately 130~s, then it decays very
rapidly as $t^{-2.0}$ until about 300~s. Finally, there is a slower decay
$\propto t^{-0.65}$. The RAPTOR, REM, and PROMPT $r'$ data (see Figure
\ref{081007fit}), and PROMPT $B$ and $i'$ data (see Figure \ref{GRB081007light
curves}) also show a similar behavior, the three $Ks$-band points also follow an
analogous decay but they are affected by large uncertainties and
therefore we disregard them in our subsequent analysis of the afterglow.                 

The temporal behavior of the early-time optical afterglow is a powerful
diagnostic of the circumburst medium density profile \citep{Pir99}, usually
modeled as a power-law, $n \propto r^{-s}$, where $n$ is the particle number
density and $r$ the distance from the burst progenitor. A homogeneous medium has
$s = 0$, while $s = 2$ represents an environment shaped by a stellar wind from
the GRB progenitor. In this scenario, the early $t^{+3}$ optical afterglow rise
could be due to the onset of either the forward shock from the outflow getting
decelerated in the interstellar medium (ISM) or of the reverse shock emission if it is
sub-relativistic \citep{Rees92,Sari98,Sari99a,Jin07}. In both cases, however, a
wind-shaped environment can be ruled out, since the optical afterglow rise
cannot be faster than $t^{1/2}$ \citep{Meszaros97,Meszaros98,Chevalier00,Jin07},
because of the rapidly decreasing circumburst density seen by the outflow.

The rapid optical decline ($\propto t^{-2}$) after the peak can only be
interpreted in the context of reverse shock emission, such as the one observed
in GRB\,990123 \citep{Akerlof99,Sari99a,Kobayashi00}. Forward shock emission can
decay so steeply only after a jet-break, which is unlikely the case at such
early time. Therefore, the early $t^{+3}$ rise is also due to the onset of the
reverse shock emission from an outflow propagating in a constant density
circumburst medium.

The \textit{Swift}/XRT light curve of GRB\,081007 can be divided into three
power-law decay stages with indices $-4.0\pm0.4$, $-0.74^{+0.03}_{-0.05}$, 
and $-1.23^{+0.11}_{-0.10}$  \citep[UK
\textit{Swift} Science Data Centre,][]{Evans09}. The last two stages are similar to the simultaneous
optical ones, which can be fitted by a $t^{-0.65}$ and a $t^{-1.25}$ decay, 
except for the last several observations, when the SN is already dominating
the flux. The SN-dominated phase can be interpreted by the sum of a power-law afterglow,
a SN template and an underlying host galaxy (see Figure \ref{081007fit}). 
The initial sharp $t^{-4}$ decay in X-rays suggests that this
emission is the tail of the prompt emission \citep[e.g.,][]{ZhMe04}. 

The X-ray and optical data between 500 and $2\times10^5$ s 
exhibit pretty much the same decay slope (see Figure \ref{081007fit}). 
The early shallow decline in both the X-ray and optical/NIR bands, roughly $t^{-0.65}$ can be
interpreted as forward shock emission for a flat electron spectral index $p
\approx 1.5$,  while the late, roughly $t^{-1.25}$, decay may be due to the jet effect as
long as the sideways expansion of the decelerating ejecta can be ignored (for
which the light curve slope before and after the break would steepen by 0.75,
close to what we observed). This interpretation is however inconsistent with the
X-ray spectrum. The time-averaged photon spectral index of second X-ray epoch (between $\sim
13$ ks) \textit{Swift}/XRT data is $2.10^{+0.15}_{-0.14}$ \citep[UK
\textit{Swift} Science Data Centre,][]{Evans09}, which suggests a normal
electron spectral index $p \approx 2.2$. We thus interpret the $t^{-0.65}$
decay, too shallow compared to model predictions, as the emission of the forward
shock with continued energy injection as ${\rm d}E/{\rm d}t \propto t^{-q}$ from
the central engine, and the $t^{-1.25}$ decay as the end of injection. 
Following \citet{Zhang06}, it is straightforward to show
that the afterglow decay index is $\alpha = [(2p-4)+(p+2)q]/4$, so the temporal
index for an energy injection index $q = 0.5$ can reproduce the data, assuming
that the optical band is above both the cooling frequency $\nu_{\rm c}$ and the
typical synchrotron radiation frequency $\nu_{\rm m}$ of the forward shock. 
When both the cooling and synchrotron radiation frequency lie redward of the optical, 
a similar intrinsic spectral index of about -1.1 is expected in X-ray and optical bands \citep{ZhMe04}, 
and is consistent with the spectral energy distribution (SED) fit by \citet{Covino13}. 
For the energy injection from a spinning-down pulsar the early energy injection rate
should have $q = 0$ \citep{Dai98a}, which is not consistent with what we
observe. Therefore, a magnetized pulsar as a central engine cannot explain the
observations.

\subsection{Interpreting the GRB\,090424 afterglow}

GRB\,090424 was also bright at early times (see Figure \ref{GRB090424light
curves}) as shown by UVOT observations, as well as $R$-band observations by TAOS
and ROTSE-III (see Figure \ref{090424fit}). The optical light curve was already
decaying at the time of the first observations and the decay index was $\approx
-1.5$, consistent with the prediction for the reverse shock emission. The
optical and X-ray light curves can be well-fitted with a simple broken power-law. 
The indices of three optical slopes are $-1.5$, $-0.85$, and 0, respectively. 
Assuming that the first phase is dominated by the reverse-shock
component, the second phase is likely dominated by the forward shock emission
that gradually overshone the fading reverse shock emission. At late times, the
emission is dominated by the host galaxy, as shown by the constant luminosity.
The best fit to the X-ray emission is four phases with decay indices $-1.29^{+0.06}_{-0.05}$, $-0.74^{+0.02}_{-0.03}$, $-1.12\pm0.02$, 
and $-1.42^{+0.18}_{-0.12}$, respectively \citep[UK \textit{Swift} Science Data Centre,][]{Evans09}. 

According to standard predictions of the fireball model in a wind-shaped
environment, the X-ray light curve should be shallower than the optical light
curve \citep[e.g.,][]{Pir99,Jin09}, which is not the case for our data.
The X-ray emission is likely due to the forward shock since, in most cases, the
reverse-shock emission in the X-ray band is very weak and can be ignored
\citep{Fan05,Xue09}. We thus interpret the X-ray emission as being due to the forward shock. 
Between 3000 and 3$\times10^5$s, its decay index 
is steeper than the optical by a factor about 0.25, 
a spectral break between X-ray and optical is then required. 
This is also confirmed by the SED fits to the X-ray and optical 
observations by \citet{Schady12} and \citet{Covino13}, a spectral break of 0.5 is required 
from X-ray to optical bands. With an ISM-like
constant density medium the forward shock is expected to be in the slow cooling
phase, i.e., the typical syncrotron frequency is below the cooling frequency
($\nu_{\rm m} < \nu_{\rm c}$). 
We find that $\nu_{\rm m} < \nu_{\rm opt} < \nu_{\rm c} <  \nu_{\rm X} $, 
here $\nu_{\rm opt}$ and $\nu_{\rm X}$ represent the optical and X-ray bands of the observations. 
We also notice that around 3 ks, the decay of the X-ray light curve steepens 
by about 0.25 \citep[the X-ray light curve before 3 ks can also be fitted with 
a single power law with decay index $-0.88$, see UK \textit{Swift} Science Data Centre,][]{Evans09}, 
which is expected when the cooling frequency $\nu_{\rm c}$ crosses 
the observational band ($\sim7\times10^{16}$~Hz, or about 0.3\,keV). Applying standard relations
\citep[e.g.,][]{ZhMe04}, this change can be interpreted as $-0.85$ ($3[p-1]/4$) to $-1.1$ ($[3p-2]/4$), 
with the electron power-law distribution index being $p = 2.13$. Given the time evolution of the cooling
frequency, $\nu_{\rm c} \propto t^{-1/2}$, it will have crossed the $R$ band
($\approx 4.2\times10^{14}$~Hz) at about $8\times10^7$~s. Finally, the last
steepening from $-1.1$ to $-1.4$ is likely due to a jet break. Fixing the final stage of X-ray 
decay to $-2$, the break is occurring at $\sim 10^6$~s. 

From the break time, the jet opening angle can be estimated as explained e.g., by
\citep{Sari99}:
\begin{equation}
  \theta_{\rm j} = 0.161\left(\frac{t_{\rm jet,day}}{1+z}\right)^{3/8}\left(\frac{n_0 \eta}{E_{\rm \gamma, iso, 52}}\right)^{1/8}~\mathrm{rad},
\end{equation}
where $E_{\rm \gamma,iso}$ is the isotropic-equivalent energy of the prompt
gamma-ray emission, $n_0$ is the number density of the medium in cm$^{-3}$ and
$\eta$ is the GRB efficiency, which we take to be $\sim 1$ and $\sim 0.2$ respectively. Here and
throughout this text, the convention $Q_x = Q/10^x$ has been adopted in cgs
units except for some special notations. Therefore for GRB\,090424, with $t_{\rm
jet,day} \sim 12$, we have $\theta_{\rm j} \sim 14^\circ$; while for
GRB\,081007, with $t_{\rm jet,day} > 12$ we have $\theta_{\rm j} > 20^\circ$.

\subsection{Estimating the Lorentz factor of the outflows}

For both GRB\,081007 and GRB\,090424 the prompt emission duration $T$ is shorter
than the outflow deceleration time $t_{\rm dec}$. In this case the reverse shock
crossing time $t_\times$ is comparable to the forward-shock deceleration time
\citep[for a review see][]{Meszaros06}, that is:
\begin{equation}
  t_\times = t_{\rm dec} = 10\left(\frac{E_{\rm k,53}}{n_0}\right)^{1/3}\Gamma_{2.5}^{-8/3}~{\rm s},
\end{equation}
where $E_{\rm k}=E_{\gamma,\rm iso}/\eta$.  The outflow Lorentz factor in turn
can be estimated as \citep[e.g.,][]{Molinari07}:
\begin{equation}
  \Gamma \sim 160\left[\frac{E_{\gamma, \rm iso, 53}(1+z)^{3}}{\eta_{0.2}n_0\,t_{\rm dec, 2}^{3}}\right]^{1/8}.
\end{equation}

For GRB\,081007, $z = 0.5295$, $E_{\gamma,\rm iso} \sim 1.5\times10^{51}$~erg,
and $t_{\rm dec} \sim 130$~s, yielding $\Gamma \sim 200(\eta_{0.2}n_0)^{1/8}$.
For GRB\,090424, $z = 0.544$, $E_{\gamma,\rm iso} \sim 4.3\times10^{52}$ erg,
and $t_{\rm dec}\leq 100$~s, yielding $\Gamma \geq 170(\eta_{0.2}n_0)^{1/8}$.

In both cases the initial GRB outflows are relativistic, consistent with what
has been found in previous works \citep[e.g.,][]{Molinari07,Sari99a,Xue09}.

\subsection{Constraining the magnetization of the outflows}

Based on the relative strength of forward- and reverse-shock emission, early GRB
afterglows can be classified into three categories \citep{Jin07, Zhang03}: Type
I, showing both peaks of the forward- and reverse-shock emission; Type II, where
the strong reverse-shock emission outshines the peak emission of the forward shock; 
Type III, where the reverse-shock emission is absent. The difference
between these three types is attributed to the very different magnetization
degrees of the outflow \citep[see ][for more detail]{Jin07}. Note that the
classification in types I, II and III has here a different meaning than in
\citet{Zhang07}. For GRB\,081007 and GRB\,090424, bright optical peaks from
the reverse-shock emission have been identified, and no forward-shock optical
peak emission can be detected, therefore, they are both Type II afterglows, for
which the reverse-shock region is expected to be mildly magnetized, as shown in
previous works \citep[e.g.,][]{Fan02,Zhang03,Fan05b}. In case the observer frequency 
lies above $\nu_{\rm m}(t_\times)$ but below $\nu_{\rm
c}(t_{\times})$, the ratio between the reverse-shock optical emission at its
crossing time ($F_{\rm obs}^{\rm r}$) and the forward-shock optical peak
emission can be estimated by \citep[as in][]{Jin07}:
\begin{equation}
  \frac{F_{\rm obs}^{\rm r}(t_{\times})}{F_{\rm obs}(t_{\rm p})} =
  0.08R_{\rm e}^{p-1}R_{\rm B}^{(p+1)/2}\left(\frac{t_{\rm p}}{t_{\times}}\right)^{3(p-1)/4},
\end{equation}

where $R_{\rm B}\equiv \varepsilon_{\rm B}^{\rm r}/\varepsilon_{\rm B}$,
$\varepsilon_{\rm B}^{\rm r}$ ($\varepsilon_{\rm B}$) is the fraction of the
reverse (forward) shock energy given to the magnetic field, $R_{\rm e} \equiv
\varepsilon_{\rm e}^{\rm r}/\varepsilon_{\rm e}$, and $\varepsilon_{\rm e}^{\rm
r}$ ($\varepsilon_{\rm e}$) is the fraction of the reverse (forward) shock
energy given to the electrons.

For GRB\,081007, we have $F_{\rm obs}^{\rm r}(t_\times) \sim 0.002$~Jy and
$t_\times \sim 130$~s. The underlying forward-shock emission peaks ($t_{\rm p}$)
between $t_{\times}$ and $\sim 300$~s (when it outshines the reverse-shock
emission), and the peak flux is $\sim 0.0005$~Jy. For $p\sim2.2$ we have $R_{\rm
B}\sim 10 R_{\rm e}^{-0.75}$.

For GRB\,090424, if we consider the first TAOS $R$-band observation \citep{Urata09} as the
reverse-shock peak, then the flux is 0.02~Jy at 90~s. The forward-shock emission
dominates the afterglow at 400~s, when the corresponding flux is 0.002~Jy.
Taking $p \sim 2.13$ and $t_{\rm p} \sim 400$~s, we have $R_{\rm B} \sim
10R_{\rm e}^{-0.72}$. It is possible that the reverse-shock peak is earlier and the flux is higher
or the forward-shock emission peak is earlier: so the derived $R_{\rm B}$ is only a lower limit.

The numerical fit to the multi-wavelength afterglow data of GRBs usually gives
$\varepsilon_{\rm e} \sim {\rm a~few} \times 0.1$ \citep[e.g.,][]{Panaitescu01}.
Since $\varepsilon_{\rm e}^{\rm r} < 1$, we have $R_{\rm e} \leq $ a few and
then $R_{\rm B} >$ a few. In other words the reverse-shock regions are mildly
magnetized. One straightforward speculation from this is that the optical flash
photons should have a moderate linear polarization degree, as has been detected
in GRB\,090102 \citep{Steele09}.

\subsection{Linking SN\,2008hw with homogenous circumburst medium}
\label{sec:100807ism}

GRB\,081007 is an event showing an optical onset and a clearly identified SN
associated with it. As discussed in Sect.\,\ref{sec:081007aft}, afterglow data
allow us to constrain the density profile of the circumburst medium, while the
occurrence of a SN confirms that the progenitor is a massive star. The afterglow
analysis suggests, as it is indeed fairly common for GRBs \citep[e.g.,][]{Schulze11},
that the outflow powering GRB\,081007 was propagating in a constant density
medium. On the other hand, a massive stellar progenitor is expected, 
during its final stages of evolution, to eject a dense wind shaping the surrounding
density profile, in possible contradiction with results based on afterglow
analysis. In the past, GRBs with bright optical
flashes and an associated SNe were  GRB\,021211/SN2002lt \citep{Fox03,Li03,DellaValle03},  
GRB\,050525A/SN\,2005nc \citep{Shao05,DellaValle06, Blustin06} and GRB\,080319B \citep{Racusin08,Bloom09,Tanvir10}.

It is widely speculated that strong episodes of mass loss occur before the death
of massive stars \citep[e.g.,][]{Pastorello07}, which is why GRBs are expected to
explode in wind-shaped media \citep{Dai98b,Chevalier00,Ofek13}. 
However, this does not seem  to be the case for GRB\,081007/SN\,2008hw, 
as we have shown in this work. A similar situation was
encountered with other nearby GRB/XRF associated with energetic and luminous
SNe: GRB\,030329/SN\,2003dh, XRF\,060218/SN\,2006aj \citep[e.g.,][]{Fan08},
although both the gamma-ray signatures and SNe were quite different
\citep{Stanek03, Hjorth06, Mazzali03, Campana06, Pian06, Mazzali06}. Also for
GRB\,090618, at a similar redshift $z = 0.54$, observations seem to favor a
homogeneous ISM environment over a wind environment \citep{Page11,Cano11}. 

The GRB\,081007 data suggest that the expansion and interaction of SN shells is
occurring in a medium with density profile more typical of a homogenous ($\sim
r^{0}$) ISM rather than a stellar wind ($\propto r^{-2}$) medium. One possible
explanation for this inconsistency is that mass loss occurring during the final
stages of the life of the massive progenitor proceeds ``discretely'' through
sudden ejection of blobs of matter  \citep[e.g.,][]{Pastorello07} rather than
``smoothly'' with a continuous, constant rate. If the time between blob-ejection 
is long enough, then there is time to redistribute matter inside the
circumstellar medium and to make it sufficiently homogeneous. Afterglow data,
extending up to 10 days after the trigger in the observer frame, can be used to
set a lower limit on the time interval that separates pulses. Since the external
shock propagates with almost luminal velocity, the distance covered by the
plasma is $\delta ct/(1+z)$, where $\delta = 1/{\Gamma [1 - \cos(\theta)]}$ is
the Doppler factor. If we assume an average value $\langle \Gamma \rangle \sim
10$ during this time, it will take $> 100$ years for a wind moving at 1000
km~s$^{-1}$ to cover this distance. This first order computation roughly
estimates the time needed for the medium surrounding the progenitor to change the
trend of the density profile from $\propto r^{-2}$ to $\propto r^{0}$. This time
can be used to constrain models of mass ejection in the late phases of massive-star 
evolution, which is still poorly understood.

\subsection{Origin of the ``single-pulse'' prompt emission of GRB\,081007}

The prompt emission of GRB\,081007 can be modeled as a single pulse
\citep{Markwardt08, Bissaldi08}. In some nearby under-luminous bursts such as
GRB\,980425 \citep{Pian00}, XRF\,060218 \citep{Campana06} and 
XRF\,100316D \citep{Starling11}, the prompt emission was also
characterized by a single pulse. The physical origin of the prompt emission is
still widely debated, with theories also including the shock breakout of the SN
explosion and the external forward-shock emission. For GRB\,081007, these two
models are actually disfavored. The inferred $\Gamma \sim 200$ is so high that
the ``shock breakout'' radius would be as large as $\sim
2\Gamma^{2}cT_{90}/(1+z) \sim 10^{16}$~cm, which is too large. In the case of
the external-shock model, a single-pulse-like prompt emission is in principle
possible and the duration of the prompt emission would trace the deceleration of
the outflow. However, the well-delineated peak of the optical afterglow, likely
marking the deceleration of the outflow at a time $\sim 130$~s, renders the
interpretation of the short-lived prompt emission as the external shock(s)
unlikely.

Below we discuss the possibility that GRB\,081007 was powered by the magnetic
energy dissipation of a Poynting-flux dominated outflow. Such a model is partly
motivated by the mild magnetization inferred from the optical afterglow data.

Following \citet{Usov94}, the radius at which the MHD condition breaks
down can be estimated following \citet{Fan05a}
\begin{equation}
  r_{\rm MHD}\sim2\times10^{16}L_{50}^{1/2} \sigma_1^{-1} t_{\rm v,m,-3}
  \Gamma_{2}^{-1}~{\rm cm},
\end{equation}
where $\sigma_1$ is the ratio of the magnetic energy flux to the particle energy
flux, $L$ is the total luminosity of the outflow, and $t_{v,m}$ is the minimum
variability timescale of the central engine. Beyond this radius, significant
magnetic dissipation processes are expected to happen which convert energy into
radiation. The radiation timescale is \citep{Gao06}
\begin{equation}
  \tau \sim {(1+z)r_{\rm MHD} \over 2 \Gamma^2 c} =
  33 ~{\rm s}~(1+z)L_{50}^{1/2} \sigma_1^{-1} t_{\rm v,m,-3} \Gamma_{2}^{-3},
\end{equation}
and the corresponding synchrotron radiation frequency can be estimated as
\citep{Fan05a,Gao06}
\begin{equation}
  \nu_{\rm m,MHD} \sim 6\times10^{16} \sigma_1^{3}C_{p}^{2} \Gamma_{2}
  t_{v,m,-3}(1+z)^{-1}~{\rm Hz},
\end{equation}
where $C_{p} \equiv (\varepsilon_{e}/0.5)[13(p-2)]/[3(p-1)]$, and
$\varepsilon_{e}$ is the fraction of the dissipated comoving magnetic-field
energy converted to the comoving kinetic energy of the electrons. Adopting
$L\sim 10^{51}$ erg~s$^{-1}$, $\sigma \sim 50$ and $\Gamma \sim 200$, the
prompt emission data of GRB\,081007 (including the duration as well as the peak
energy) can be well reproduced.
 
After the identification of a distinct thermal radiation component in
GRB\,090902B \citep[e.g.,][]{Ryde10,Zhang11}, the photospheric radiation model
has attracted wide attention. In such a scenario, the prompt gamma-rays are
produced by the significantly modified quasi-thermal radiation from the
photosphere of the outflow or from sites with an optical depth of $\sim 10$
\citep[][and references therein]{Beloborodov13}. For GRB\,081007, such an
origin can not be ruled out. With future $\gamma-$ray polarimetry data
\citep[see][for preliminary results]{Gotz09,Yonetoku11} we may be able to
distinguish  between the global-magnetic-energy-dissipation model and the
photospheric-radiation model, since the former usually predicts a moderate or
high linear polarization while the latter usually does not. One exception is
that in a specific photosphere model moderate linear polarization is possible if
our line of sight happens to be at the edge of the ejecta \citep{Fan09}.

\subsection{Host galaxy parameters}
\label{diagnosehost}

The apparent magnitudes of  host galaxy of GRB\,090424 are $ r = 22.07\pm0.12$ and
$i = 21.82\pm0.12$, so the absolute magnitudes are approximately $M_B =
-20.47\pm0.12$ and $M_V = -20.69\pm0.12$, considering the Galactic extinction
$E(B-V)=0.025$. These figures are close to the values for our Galaxy, and are
brighter than for most GRB hosts (see e.g., \citealt{Hjorth12}).

The metallicity of a galaxy can be derived from the ratio of different emission
lines in its spectrum. Several methods have been studied and adopted \citep[for
a recent review, see][]{Kewley08}. In our case, we used the N2 and O3N2 indices,
as recalibrated by \citet{Pettini04}. We find a metallicity
$12+\log(\mathrm{O/H})$ of about 8.39 and 8.43 using the N2 and O3N2 methods,
respectively. These two methods are also weakly affected by intrinsic or
Galactic extinction.

The observed fluxes of the H$\alpha$ and H$\beta$ emission lines are
$151.64\pm8.45$ and $20.49\pm3.04\times10^{-18}$ erg s$^{-1}$ cm$^{-2}$. To
estimate the SFR from the H$\alpha$ or H$\beta$ line, we followed the relations used
in \citet{Savaglio09}: SFR$_{\rm H\alpha} = 4.39\times10^{-42}L_{\rm H\alpha}\,
M_\odot\,{\rm yr}^{-1}$ and SFR$_{\rm H\beta} = 12.6\times10^{-42}L_{\rm
H\beta}\,M_\odot\,{\rm yr}^{-1}$. Correcting for Galactic extinction, the lower limit 
(since the host galaxy extinction is not corrected) of
SFR is SFR$_{\rm H \alpha} = 0.80\, M_\odot\,{\rm yr}^{-1}$ or SFR$_{\rm H\beta}
= 0.32\,M_\odot\,{\rm yr}^{-1}$.

\section{Summary}

In this paper we have presented and interpreted multi-wavelength observations of
GRB\,081007 and GRB\,090424, and we  summarize here the results.

\textit{i)} The early stages of both afterglows are characterized by a bright
optical/NIR component, which we interpret as the reverse-shock emission.

\textit{ii)} The late-time afterglow of GRB\,081007 is dominated by a SN
component (SN\,2008hw) similar to SN\,1998bw near maximum light. The presence of
a SN associated with the GRB clearly suggests that this burst originated from a
massive star that should have shaped its circumburst environment with wind.
On the other hand, the afterglow data can only be interpreted assuming a
surrounding ISM-like medium, characterized by a constant density profile. A
process to make the circumburst medium around the progenitor star homogeneous
has likely been effective.

\textit{iii)} The entire set of the afterglow data of GRB\,081007 can be
interpreted within the forward- and reverse-shock model, consider a long-lasting
energy injection following the law $\mathrm{d}E/\mathrm{d}t \propto t^{-0.5}$.

\textit{iv)} The initial Lorentz factor of GRB\,081007 outflow is estimated by
the afterglow data to be $\Gamma \sim 200$, which makes the interpretation of
its single-pulse prompt emission in terms of both external forward shock and
shock breakout unlikely. The identification of a reverse-shock emission
component peaking at $\sim 130$~s after trigger rules out the possibility that
the short-lived prompt emission was due to external-shock emission. The
absence of the peak of forward-shock optical emission strongly suggests that the
reverse-shock regions should be mildly magnetized. We therefore suggest that the
prompt emission, characterized by a single pulse, may be due to the magnetic
energy dissipation of a Poynting-flux dominated outflow or to a dissipative
photosphere.  

\textit{v)} For GRB\,090424, we set a lower limit on the initial Lorentz factor
of the outflow of $\Gamma > 170$. Unlike GRB\,081007, we did not detect the SN
component in the afterglow, likely due to the considerably bright host galaxy,
roughly comparable to the Milky Way. The bright initial optical/NIR afterglow
has also been attributed to emission from a mildly-magnetized reverse shock. The
late time X-ray and optical data are consistent with the forward-shock model and
the surrounding medium is also found to be ISM-like.

All these results demonstrate that multi-band afterglow data, in particular with
very early observations, are a necessary and valuable tool to better understand
GRB physics. Significant progress is expected in the near future as more
data will be collected.


\section*{Acknowledgments}
We acknowledge the support from the ASI grants I/011/07/0 and I/088/06/0, 
the INAF PRIN 2009 and 2011, and the MIUR PRIN 2009ERC3HT. 
We thank the Paranal Science Operations staff, and in particular T. Rivinius, P. Lynam,  S. Brillant, 
F.J. Selman, T. Szeifert, L. Schmidtobreick, A. Smette, A. Ahumada, K. O'Brien, C. Ledoux for effectively carrying out our service-mode observations. 
We wish to acknowledge the anonymous referee who has significantly improved the presentation and the discussion of the data. 
The Dark Cosmology Centre is funded by the DNRF. 
J.P.U.F. acknowledges support from the ERC-StG grant EGGS-278202. 
The RAPTOR project is supported by the DOE sponsored LDRD program at LANL. 
The Liverpool Telescope is operated by Liverpool John Moores University at the Observatorio del Roque de los Muchachos of the Instituto de Astrof\'{i}sica de Canarias. 
The Faulkes Telescopes are owned by Las Cumbres Observatory. CGM acknowledges support from the Royal Society.
Data collected with the Gran Telescopio Canarias (GTC), 
installed in the Spanish Observatorio del Roque de los Muchachos of the Instituto de Astrof\'isica de Canarias, on the island of La Palma. 
Based on observations collected with the 3.5m telescope of the German-Spanish Astronomical Centre, Calar Alto, Spain, 
operated jointly by the Max-Plank-Institut f\"ur Astronomie (MPIA), Heidelberg, and the Spanish National Commission for Astronomy. 
This study was carried out in the framework of the Unidad Asociada IAA-CSIC at the group of planetary science of ETSI-UPV/EHU 
and supported by the Ikerbasque Foundation for Science. 
The research of J.G., A.J.C.T and R.S.R. is supported by the Spanish programmes AYA2008-03467/ESP, 
AYA2012-39727-C03-01 and AYA2009-14000-C03-01. 
Z.P.J. thanks Dr. Yi-Zhong Fan for stimulating discussion. 
Z.P.J. was supported by the National Natural Science Foundation of China under the grants 11073057 and 11103084.

\section*{Appendix: Table for observations\footnote{The following material only appears online}}

\begin{table}
\caption{Optical observations of GRB\,081007.}

\begin{minipage}[t]{\textwidth}
\begin{minipage}[t]{0.5\textwidth}
\begin{tabular}{r r r r r r}
\hline
Instrument & Filter & Mid-time & Exp-time & Mag & Error \\
 &  & (min) & (min) & (AB) & \\
\hline
REM & $r'$ & 2.94 & 2.50 & 16.35 & 0.06\\
 & &5.67 & 0.01 & 17.14 & 0.24\\
 & &6.32 & 0.01 & 17.41 & 0.31\\
 & &6.98 & 0.01 & 17.11 & 0.24\\
 & &7.64 & 0.01 & 17.20 & 0.26\\
 & &8.29 & 0.01 & 17.39 & 0.31\\
 & &8.94 & 0.01 & 17.28 & 0.32\\
 
 & $H$ &1.57 & 0.08 & 15.41 & 0.29  \\
 & &1.82 & 0.08 & 15.00 & 0.14  \\
 & &2.08 & 0.08 & 14.47 & 0.20   \\
 & &2.38 & 0.08 & 14.80 & 0.12 \\
 & &2.64 & 0.08 & 15.09 & 0.12 \\
 & &2.88 & 0.08 & 15.34 & 0.18 \\
 & &3.40 & 0.08 & 15.43 & 0.19 \\
 & &4.21 & 0.50 & 16.11 & 0.14  \\
 & &5.53 & 0.52 & 16.20 & 0.14   \\
 & &10.68 & 0.50 & 16.68 & 0.24  \\
 & &16.51 & 0.85 & 17.37 & 0.23  \\
 & &25.67 & 1.17 & 17.42 & 0.21  \\
 & &50.85 & 1.18 & 17.90 & 0.31  \\

 & $K_s$ &7.68 & 0.39 & 16.80 & 0.18 \\
 & &12.13 & 0.59 & 17.21 & 0.19 \\
 & &18.85 & 1.01 & 17.57 & 0.17 \\
 
RAPTOR & $r'$ &1.51 & 0.08 & 17.06 & 0.04 \\
 & &1.66 & 0.08 & 16.64 & 0.03 \\
 & &1.90 & 0.17 & 16.07 & 0.02 \\
 & &2.11 & 0.17 & 15.93 & 0.02 \\
 & &2.33 & 0.17 & 15.84 & 0.02 \\
 & &2.55 & 0.17 & 15.90 & 0.02 \\
 & &2.76 & 0.17 & 15.92 & 0.02 \\
 & &2.98 & 0.17 & 16.14 & 0.03 \\
 & &3.20 & 0.17 & 16.10 & 0.03 \\
 & &3.42 & 0.17 & 16.47 & 0.04 \\
 & &3.63 & 0.17 & 16.51 & 0.04 \\
 & &3.84 & 0.17 & 16.50 & 0.04 \\
 & &4.05 & 0.17 & 16.65 & 0.04 \\
 & &4.26 & 0.17 & 16.93 & 0.05 \\
 & &4.48 & 0.17 & 16.75 & 0.05 \\
 & &4.69 & 0.17 & 16.79 & 0.05 \\
 & &4.90 & 0.17 & 16.73 & 0.04 \\
 & &5.12 & 0.17 & 16.99 & 0.06 \\
 & &5.33 & 0.17 & 16.97 & 0.06 \\
 & &5.54 & 0.17 & 17.34 & 0.08 \\
 & &6.16 & 0.97 & 17.14 & 0.03 \\
 & &7.64 & 1.76 & 17.43 & 0.07 \\
 & &9.53 & 1.77 & 17.93 & 0.11 \\
 & &11.42 & 1.76 & 17.90 & 0.11 \\
 & &13.94 & 3.01 & 18.30 & 0.12 \\
 & &17.08 & 3.00 & 18.61 & 0.16 \\
 & &21.80 & 6.18 & 18.75 & 0.13 \\

UVOT & $u$ &11.37 & 0.33 & 18.40 & 0.35 \\
 & &13.87 & 0.33 & 18.14 & 0.29 \\
 & &418.62 & 4.92 & 20.34 & 0.39 \\
 & &423.68 & 4.92 & 20.15 & 0.33 \\
 & &993.28 & 6.95 & 20.40 & 0.35 \\

 & $b$ &14.21 & 0.16 & 18.19 & 0.32 \\

 & $v$ &3.78 & 0.40 & 16.77 & 0.16 \\
 & &4.19 & 0.42 & 16.87 & 0.17 \\
 & &4.61 & 0.42 & 17.97 & 0.35 \\
 & &5.03 & 0.42 & 17.31 & 0.22 \\
 & &5.44 & 0.42 & 17.35 & 0.23 \\
 & &5.86 & 0.42 & 18.21 & 0.41 \\
 & &6.28 & 0.42 & 17.60 & 0.27 \\
 & &6.69 & 0.42 & 17.54 & 0.25 \\
 & &7.11 & 0.42 & 17.42 & 0.24 \\
 & &7.53 & 0.42 & 17.85 & 0.31 \\
 
\hline
\end{tabular}
\end{minipage}
\begin{minipage}[t]{0.5\textwidth}
\begin{tabular}{r r r r r r}
\hline
Instrument & Filter & Mid-time & Exp-time & Mag & Error  \\
 &  & (min) & (min) & (AB) & \\
\hline

UVOT & $v$ &7.94 & 0.42 & 17.55 & 0.26 \\
 & &8.36 & 0.42 & 17.82 & 0.32 \\
 & &9.19 & 0.42 & 17.67 & 0.29 \\
 & &9.61 & 0.42 & 18.11 & 0.38 \\
 & &17.32 & 1.67 & 18.63 & 0.29 \\
 & &18.98 & 1.67 & 18.96 & 0.37 \\
 & &20.65 & 1.67 & 18.45 & 0.25 \\
 
 & White & 2.02 & 0.42 & 17.08 & 0.06  \\
 & &2.43 & 0.42 & 17.00 & 0.06  \\
 & &2.85 & 0.42 & 17.17 & 0.07  \\
 & &3.27 & 0.41 & 17.37 & 0.07  \\
 & &12.01 & 0.16 & 18.81 & 0.28  \\
 & &322.75 & 4.92 & 20.17 & 0.15  \\
 & &327.62 & 4.57 & 20.27 & 0.17  \\
 & &1000.44 & 6.95 & 21.14 & 0.28  \\
 & &1753.79 & 12.97 & 21.44 & 0.27 \\
 
PROMPT & $B$ &1.86 & 0.33 & 16.77 & 0.16 \\
 & &2.34 & 0.33 & 16.48 & 0.17 \\
 & &2.80 & 0.33 & 16.94 & 0.21 \\
 & &6.68 & 3.42 & 18.10 & 0.13 \\
 & &19.88 & 8.28 & 18.46 & 0.11 \\
 & &47.06 & 17.34 & 19.24 & 0.15 \\
 & &118.19 & 52.34 & 20.13 & 0.27 \\
 
 & $V$ & 2.41 & 0.17 & 16.51 & 0.23 \\ 
 & &  3.27 & 0.17 & 16.71 & 0.23 \\ 
 & &  4.50 & 0.67 & 17.15 & 0.11 \\ 
 & &  6.18 & 0.67 & 17.66 & 0.18 \\ 
 & &  8.57 & 1.33 & 17.85 & 0.12 \\ 
 & &11.61 & 1.33 & 17.89 & 0.12 \\ 
 & &14.64 & 1.33 & 17.97 & 0.14 \\ 
 & &16.13 & 1.33 & 17.91 & 0.14 \\ 
 & &20.65 & 1.33 & 18.06 & 0.14 \\ 
 & &22.15 & 1.33 & 18.26 & 0.15 \\ 
 & &26.70 & 1.33 & 18.41 & 0.20 \\ 
 & &35.73 & 1.33 & 18.21 & 0.14 \\ 
 & &37.21 & 1.33 & 18.54 & 0.18 \\ 
 & &38.68 & 1.33 & 18.65 & 0.21 \\ 
 & &49.20 & 1.33 & 18.69 & 0.23 \\ 
 & &64.40 & 1.33 & 18.94 & 0.13 \\ 
 
& $r'$ & 2.33 & 0.75 & 16.03 & 0.07\\ 
 & &6.27 & 2.67 & 16.80 & 0.05\\ 
 & &39.13 & 28.00 & 18.52 & 0.13\\ 
 & &902.94 & 176.00 & 20.65 & 0.15\\ 

 & $i'$ &1.27 & 0.33 & 17.43 & 0.10 \\ 
 & &2.35 & 0.67 & 15.42 & 0.05 \\ 
 & &3.09 & 0.50 & 15.71 & 0.05 \\ 
 & &4.03 & 0.67 & 16.22 & 0.20 \\ 
 & &5.37 & 1.33 & 16.57 & 0.05 \\ 
 & &7.48 & 3.33 & 16.97 & 0.03 \\ 
 & &12.27 & 4.00 & 17.54 & 0.03 \\ 
 & &16.78 & 4.00 & 17.68 & 0.03 \\ 
 & &21.32 & 4.00 & 17.96 & 0.05 \\ 
 & &25.83 & 4.00 & 18.25 & 0.05 \\ 
 & &30.38 & 4.00 & 18.23 & 0.05 \\ 
 & &34.87 & 4.00 & 18.17 & 0.05 \\ 
 & &39.27 & 4.00 & 18.27 & 0.05 \\ 
 & &43.87 & 4.00 & 18.46 & 0.06 \\ 
 & &48.37 & 4.00 & 18.50 & 0.05 \\ 
 & &55.97 & 8.00 & 18.42 & 0.07 \\ 
 & &64.40 & 8.00 & 19.10 & 0.08 \\ 
 & &76.85 & 13.33 & 18.64 & 0.05 \\ 
 & &92.02 & 13.33 & 18.77 & 0.05 \\ 
 & &108.95 & 16.00 & 18.99 & 0.07 \\ 
 & &126.90 & 16.00 & 19.37 & 0.08 \\ 
 & &187.87 & 49.33 & 19.14 & 0.09 \\ 
 & &1206.90 & 66.67 & 21.38 & 0.12 \\ 
 
\hline
\end{tabular}
\end{minipage}
\end{minipage}
\label{081007data}
\end{table}

\begin{table}
\caption{Optical observations of GRB\,081007 (continued). }

\begin{minipage}[t]{\textwidth}
\begin{minipage}[t]{0.5\textwidth}
\begin{tabular}{r r r r r r}
\hline
Instrument & Filter & Mid-time & Exp-time & Mag & Error  \\
 &  & (min) & (min) & (AB) & \\
\hline

FTN & $B$ &18.03 & 0.17 & 19.06 & 0.31 \\
 & &22.93 & 0.50 & 19.00 & 0.22 \\
 & &27.33 & 1.00 & 19.49 & 0.21 \\
 & &33.60 & 2.00 & 19.77 & 0.21 \\
 & &42.80 & 3.00 & 19.83 & 0.18 \\
 & &53.95 & 2.00 & 19.73 & 0.19 \\
 & &64.13 & 3.00 & 19.98 & 0.18 \\
 & &75.30 & 4.00 & 20.01 & 0.18 \\
 & &109.78 & 10.00 & 20.35 & 0.15 \\
 & &236.48 & 10.00 & 20.77 & 0.17 \\
 & &1584.32 & 30.00 & 22.66 & 0.34 \\

 & $V$ &20.38 & 0.17 & 18.59 & 0.14 \\
 & &84.25 & 10.00 & 19.67 & 0.04 \\
 & &145.95 & 10.00 & 19.69 & 0.08 \\
 & &224.45 & 10.00 & 20.15 & 0.06 \\
 & &1642.58 & 10.00 & 21.73 & 0.26 \\
 & &1668.58 & 15.00 & 22.51 & 0.49 \\

 & $r'$ &17.49 & 0.50 & 18.21 & 0.05 \\
 & &24.27 & 0.50 & 18.50 & 0.06 \\
 & &29.15 & 1.00 & 18.54 & 0.04 \\
 & &36.42 & 2.00 & 18.75 & 0.05 \\
 & &46.57 & 3.00 & 18.87 & 0.03 \\
 & &56.75 & 2.00 & 19.10 & 0.04 \\
 & &66.90 & 3.00 & 19.13 & 0.06 \\
 & &133.94 & 10.00 & 19.71 & 0.05 \\
 & &203.04 & 10.00 & 19.95 & 0.05 \\
 & &260.67 & 10.00 & 19.90 & 0.05 \\
 & &1581.47 & 30.00 & 21.60 & 0.16 \\
 & &2981.45 & 60.00 & 22.74 & 0.37 \\

 & $i'$ &21.55 & 0.17 & 17.90 & 0.08 \\
 & &25.67 & 0.50 & 18.06 & 0.05 \\
 & &31.13 & 1.00 & 18.25 & 0.04 \\
 & &39.38 & 2.00 & 18.40 & 0.04 \\
 & &51.58 & 3.00 & 18.62 & 0.03 \\
 & &59.72 & 2.00 & 18.71 & 0.04 \\
 & &71.88 & 3.00 & 18.87 & 0.04 \\
 & &121.88 & 10.00 & 19.26 & 0.07 \\
 & &190.98 & 10.00 & 19.61 & 0.05 \\
 & &248.62 & 10.00 & 19.54 & 0.05 \\
 & &1604.16 & 35.00 & 21.62 & 0.15 \\
 & &3015.03 & 60.00 & 22.56 & 0.45 \\

FTS & $r'$ &1710.34 & 60.00 & 21.29 & 0.18 \\
 & &3395.84 & 90.00 & 22.09 & 0.38 \\

 & $i'$ &1712.61 & 30.00 & 21.41 & 0.19 \\

VLT & $R_c$ & 24163.20 & 13.70 & 23.65 & 0.07 \\
 & &38563.20 &2.00 &23.79 &0.13 \\
 & &60264.00 &6.00 &24.14 &0.10 \\
 & &73152.00 &6.00 &24.39 &0.11 \\
 & &87566.40 &6.00 &24.37 &0.18 \\
 & &97660.80 &12.00 &24.44 &0.12 \\
 & &125712.00 &27.00 &$>$24.67 &  \\
 
 & $I_c$ &60278.40 &6.00 &23.18 &0.10 \\
 & &73166.40 &	6.00 &23.67 &0.11\\ 
 & &87566.40 &6.00 &24.12 &0.22 \\ 
 & &97660.80 &12.00 &23.93 &0.16\\ 
 & &126475.20 &21.00 &24.29 &0.20\\ 
 
Gemini & $g'$ &24156.00 & 5.00 & 24.91 & 0.19 \\

 & $r'$ &24173.00 & 5.00 & 23.89 & 0.08 \\

 & $i'$ &24179.00 & 5.00 & 23.58 & 0.07 \\

 & $z'$ &24186.00 & 5.00 & 23.13 & 0.09 \\

\hline
\end{tabular}
\end{minipage}
\end{minipage}
\tablecomments{Galactic extinction has not been removed.}
\end{table}

\begin{table}
\caption{Optical observations of GRB\,090424. }

\begin{minipage}[t]{\textwidth}
\begin{minipage}[t]{0.5\textwidth}
\begin{tabular}{r r r r r r}
\hline
Instrument & Filter & Mid-time & Exp-time & Mag & Error  \\
 &  & (min) & (min) & (AB) & \\
\hline
UVOT & UVW2 &13.22 & 0.32 & 17.54 & 0.28 \\
 & &19.94 & 0.32 & 17.66 & 0.29 \\
 & &133.50 & 2.02 & 19.77 & 0.38 \\

 & UVM2 & 11.17 & 0.32 & 17.39 & 0.34 \\
 & & 320.28 & 13.80 & 20.04 & 0.22 \\
 & & 603.99 & 14.76 & 20.59 & 0.31 \\

 & UVW1 & 11.58 & 0.33 & 16.77 & 0.19 \\
 & &18.30 & 0.32 & 17.93 & 0.34 \\
 & &21.17 & 0.32 & 17.49 & 0.28 \\
 & &24.04 & 0.32 & 17.69 & 0.31 \\
 & &26.94 & 0.33 & 18.49 & 0.48 \\
 & &29.82 & 0.32 & 18.20 & 0.41 \\
 & &32.71 & 0.33 & 18.66 & 0.52 \\
 & &35.58 & 0.33 & 17.56 & 0.28 \\
 & &94.17 & 3.28 & 18.79 & 0.19 \\
 & &417.97 & 11.20 & 20.52 & 0.33 \\

 & $u$ &5.15 & 0.17 & 16.00 & 0.13 \\
 & &5.32 & 0.17 & 16.29 & 0.15 \\
 & &5.48 & 0.17 & 16.25 & 0.15 \\
 & &5.65 & 0.17 & 16.23 & 0.14 \\
 & &5.82 & 0.17 & 16.36 & 0.15 \\
 & &5.98 & 0.17 & 16.35 & 0.15 \\
 & &6.15 & 0.17 & 16.73 & 0.19 \\
 & &6.32 & 0.17 & 16.46 & 0.16 \\
 & &6.48 & 0.17 & 16.47 & 0.16 \\
 & &6.65 & 0.17 & 16.75 & 0.19 \\
 & &6.82 & 0.17 & 16.38 & 0.15 \\
 & &6.98 & 0.17 & 16.87 & 0.20 \\
 & &7.15 & 0.17 & 16.76 & 0.19 \\
 & &7.32 & 0.17 & 16.63 & 0.18 \\
 & &7.48 & 0.17 & 17.05 & 0.22 \\
 & &7.65 & 0.17 & 16.73 & 0.18 \\
 & &7.82 & 0.17 & 16.68 & 0.18 \\
 & &7.98 & 0.17 & 16.55 & 0.17 \\
 & &8.15 & 0.17 & 17.00 & 0.22 \\
 & &8.32 & 0.17 & 16.91 & 0.20 \\
 & &8.48 & 0.17 & 16.83 & 0.19 \\
 & &8.65 & 0.17 & 16.63 & 0.17 \\
 & &8.82 & 0.17 & 17.29 & 0.25 \\
 & &8.98 & 0.17 & 16.78 & 0.19 \\
 & &9.15 & 0.16 & 16.84 & 0.20 \\
 & &11.82 & 0.09 & 17.07 & 0.30 \\
 & &11.98 & 0.17 & 17.19 & 0.23 \\
 & &12.15 & 0.08 & 17.30 & 0.37 \\
 & &18.70 & 0.33 & 17.41 & 0.19 \\
 & &24.45 & 0.33 & 17.53 & 0.20 \\
 & &27.35 & 0.33 & 17.73 & 0.23 \\
 & &33.12 & 0.32 & 18.65 & 0.42 \\
 & &35.99 & 0.32 & 18.04 & 0.28 \\
 & &96.14 & 0.44 & 18.01 & 0.24 \\
 & &103.97 & 3.28 & 19.12 & 0.19 \\
 & &500.30 & 4.92 & 20.01 & 0.31 \\
 & &505.37 & 4.92 & 20.23 & 0.37 \\
 & &510.43 & 4.92 & 19.86 & 0.27 \\

 & $b$ &9.51 & 0.33 & 17.32 & 0.14 \\
 & &12.39 & 0.33 & 17.46 & 0.14 \\
 & &19.12 & 0.32 & 17.84 & 0.18 \\
 & &21.98 & 0.32 & 18.06 & 0.21 \\
 & &24.89 & 0.33 & 18.30 & 0.24 \\
 & &27.77 & 0.33 & 18.56 & 0.29 \\
 & &30.64 & 0.33 & 18.36 & 0.26 \\
 & &33.53 & 0.32 & 18.75 & 0.34 \\
 & &36.40 & 0.32 & 18.07 & 0.21 \\
 & &107.40 & 3.28 & 19.79 & 0.24 \\
 & &229.69 & 2.41 & 20.13 & 0.37 \\
 & &515.53 & 4.92 & 20.29 & 0.30 \\
 & &519.06 & 1.92 & 20.80 & 0.74 \\

\hline
\end{tabular}
\end{minipage}
\begin{minipage}[t]{0.5\textwidth}
\begin{tabular}{r r r r r r}
\hline
Instrument & Filter & Mid-time & Exp-time & Mag & Error  \\
 &  & (min) & (min) & (AB) & \\
\hline

UVOT & $v$ &1.33 & 0.16 & 13.82 & 0.07 \\
 & &10.74 & 0.33 & 16.88 & 0.19 \\
 & &13.64 & 0.33 & 17.19 & 0.24 \\
 & &17.48 & 0.33 & 17.65 & 0.31 \\
 & &20.35 & 0.33 & 17.97 & 0.39 \\
 & &23.21 & 0.33 & 17.89 & 0.37 \\
 & &26.11 & 0.33 & 17.79 & 0.35 \\
 & &29.00 & 0.33 & 18.75 & 0.72 \\
 & &31.89 & 0.32 & 17.63 & 0.32 \\
 & &34.76 & 0.32 & 18.35 & 0.53 \\
 & &37.66 & 0.32 & 17.58 & 0.30 \\
 & &200.71 & 4.92 & 19.55 & 0.35 \\
 & &203.53 & 0.52 & 19.77 & 1.33 \\

 & White &1.62 & 0.17 & 15.03 & 0.05 \\
 & &1.78 & 0.17 & 15.39 & 0.05 \\
 & &1.95 & 0.17 & 15.27 & 0.05 \\
 & &2.12 & 0.17 & 15.46 & 0.05 \\
 & &2.28 & 0.17 & 15.57 & 0.06 \\
 & &2.45 & 0.17 & 15.43 & 0.05 \\
 & &2.62 & 0.17 & 15.55 & 0.06 \\
 & &2.78 & 0.17 & 15.66 & 0.06 \\
 & &2.95 & 0.17 & 15.70 & 0.06 \\
 & &3.12 & 0.17 & 15.77 & 0.06 \\
 & &3.28 & 0.17 & 15.87 & 0.06 \\
 & &3.45 & 0.17 & 15.99 & 0.06 \\
 & &3.62 & 0.17 & 16.11 & 0.06 \\
 & &3.78 & 0.17 & 16.06 & 0.06 \\
 & &3.95 & 0.16 & 16.23 & 0.07 \\
 & &9.78 & 0.12 & 17.26 & 0.12 \\
 & &9.95 & 0.17 & 17.55 & 0.13 \\
 & &10.12 & 0.04 & 17.77 & 0.29 \\
 & &12.62 & 0.07 & 17.51 & 0.19 \\
 & &12.78 & 0.17 & 17.66 & 0.13 \\
 & &12.95 & 0.10 & 18.00 & 0.21 \\
 & &19.45 & 0.17 & 18.37 & 0.21 \\
 & &19.62 & 0.15 & 18.11 & 0.18 \\
 & &22.28 & 0.14 & 18.28 & 0.21 \\
 & &22.45 & 0.17 & 18.07 & 0.17 \\
 & &25.28 & 0.33 & 18.31 & 0.14 \\
 & &28.17 & 0.32 & 18.06 & 0.13 \\
 & &31.06 & 0.33 & 18.14 & 0.13 \\
 & &33.93 & 0.32 & 18.29 & 0.15 \\
 & &36.81 & 0.33 & 18.16 & 0.13 \\
 & &110.47 & 2.61 & 19.26 & 0.11 \\

LT & $r'$ & 453.22 & 30.00 & 19.67 & 0.02 \\
 & & 573.28 & 30.00 & 19.91 & 0.08 \\

 & $i'$ & 499.33 & 30.00 & 19.66 & 0.04 \\

FTN & $r'$ & 25435.85 & 30.00 & 21.86 & 0.11 \\
 & & 480809.88 & 30.00 & 21.65 & 0.11 \\
 & & 482086.00 & 30.00 & 22.03 & 0.15 \\
 & & 483666.35 & 30.00 & 21.78 & 0.15 \\

VLT & $r'$ & 9360.01 & 6.00 & 22.00 & 0.10 \\
 & & 26619.71 & 6.00 & 22.08 & 0.15 \\
 & & 38151.23 & 6.00 & 22.13 & 0.08 \\
 & & 71316.72 & 6.00 & 21.90 & 0.35 \\
 & & 104267.52 & 3.00 & 22.50 & 0.21 \\

 & $i'$ & 9344.68 & 4.00 & 21.65 & 0.08 \\
 & & 9349.57 & 4.00 & 21.56 & 0.08 \\
 & & 9354.13 & 4.00 & 21.66 & 0.08 \\
 & & 26627.66 & 4.00 & 21.80 & 0.10 \\
 & & 26634.29 & 4.00 & 21.79 & 0.09 \\
 & & 38156.93 & 4.00 & 21.78 & 0.06 \\
 & & 38161.57 & 4.00 & 21.77 & 0.07 \\
 & & 71322.34 & 4.00 & 21.88 & 0.14 \\
 & & 71326.81 & 4.00 & 21.85 & 0.16 \\
 & & 104238.34 & 4.00 & 21.86 & 0.13 \\
 & & 104247.08 & 4.00 & 21.85 & 0.13 \\
 
CAHA & $J$ & 1589029.99 & 212.00 & 21.60 & 0.11 \\
 & $H$ & 1589080.06 & 47.00 & 21.21& 0.32 \\

\hline
\end{tabular}
\end{minipage}
\end{minipage}
\tablecomments{Galactic extinction has not been removed.}
\label{090424data}
\end{table}


\begin{thebibliography}{99}
\bibitem[\protect\citeauthoryear{Akerlof et al.}{1999}]{Akerlof99}Akerlof, C., Balsano, R., Barthelmy, S., et al. 1999, \nat, 398, 400
\bibitem[\protect\citeauthoryear{Baumgartner et al.}{2008}]{Baumgartner08} Baumgartner, W. H., Cummings, J. R., Evans, P. A., et al. 2008, GCN, 8330
\bibitem[\protect\citeauthoryear{Beloborodov}{2013}]{Beloborodov13} Beloborodov, A. M. 2013, \apj, 764, 157
\bibitem[\protect\citeauthoryear{Berger et al.}{2011}]{Berger11} Berger, E., Chornock, R., Holmes, T. R., et al. 2011, ApJ, 743, 204
\bibitem[\protect\citeauthoryear{Berger et al.}{2008}]{Berger08} Berger, E., Fox, D. B., Cucchiara, A., \& Cenko, S. B. 2008, GCN, 8335
\bibitem[\protect\citeauthoryear{Bissaldi et al.}{2008}]{Bissaldi08} Bissaldi, E., McBreen, S., \& Connaughton, V. 2008, GCN, 8369
\bibitem[\protect\citeauthoryear{Bloom et al.}{2009}]{Bloom09} Bloom, J. S., Perley, D. A., Li, W., et al. 2009, \apj, 691, 723
\bibitem[\protect\citeauthoryear{Blustin et al.}{2006}]{Blustin06} Blustin, A. J., Band, D., Barthelmy, S., et al. 2006, \apj, 637, 901
\bibitem[\protect\citeauthoryear{Campana et al.}{2006}]{Campana06} Campana, S., Mangano, V., Blustin, A. J., et al. 2006, \nat, 442, 1008
\bibitem[\protect\citeauthoryear{Cannizzo et al.}{2009}]{Cannizzo09} Cannizzo, J. K., Barthelmy, S. D., Beardmore, A. P., et al. 2009, GCN, 9223
\bibitem[\protect\citeauthoryear{Cano et al.}{2011}]{Cano11} Cano, Z., Bersier, D., Guidorzi, C., et al. 2011, \apj, 740, 41
\bibitem[\protect\citeauthoryear{Chevalier \& Li}{2000}]{Chevalier00} Chevalier, R. A., \& Li, Z. Y. 2000, \apj, 536, 195
\bibitem[\protect\citeauthoryear{Chornock et al.}{2009}]{Chornock09} Chornock, R., Perley, D. A., Cenko, S. B., \& Bloom, J. S. 2009, GCN, 9243
\bibitem[\protect\citeauthoryear{Cobb}{2008}]{Cobb08} Cobb, B. E. 2008, GCN, 8339
\bibitem[\protect\citeauthoryear{Cobb}{2009}]{Cobb09} Cobb, B. E. 2009, GCN, 9313 
\bibitem[\protect\citeauthoryear{Colgate}{1968}]{Colgate68} Colgate, S. A. 1968, CaJPh, 46, 476
\bibitem[\protect\citeauthoryear{Covino et al.}{2008}]{Covino08} Covino, S., Antonelli, L. A., Calzoletti, L., et al. 2008, GCN, 8331
\bibitem[\protect\citeauthoryear{Covino et al.}{2013}]{Covino13} Covino, S., Melandri, A., Salvaterra, R., et al. 2013, \mnras, 432, 1231
\bibitem[\protect\citeauthoryear{Dai \& Lu}{1998a}]{Dai98a} Dai, Z. G., \& Lu, T. 1998a, \aap, 333, L87
\bibitem[\protect\citeauthoryear{Dai \& Lu}{1998b}]{Dai98b} Dai, Z. G., \& Lu, T. 1998b, \mnras, 298, 87
\bibitem[\protect\citeauthoryear{de Ugarte Postigo et al.}{2011}]{deUgartePostigo11} de Ugarte Postigo, A., Horv{\'a}th, I., Veres, P., et al. 2011, \aap, 525, 109
\bibitem[\protect\citeauthoryear{Della Valle}{2011}]{DellaValle11} Della Valle, M. 2011, IJMPD, 20, 1745
\bibitem[\protect\citeauthoryear{Della Valle et al.}{2008}]{DellaValle08} Della Valle, M., Benetti, S., Mazzali, P., et al. 2008, CBET, 1602
\bibitem[\protect\citeauthoryear{Della Valle et al.}{2003}]{DellaValle03} Della Valle, M., Malesani, D., Benetti, S., et al. 2003, \aap, 406, L33
\bibitem[\protect\citeauthoryear{Della Valle et al.}{2006}]{DellaValle06} Della Valle, M., Malesani, D., Bloom, J. S., et al. 2006, \apj, 642, L103
\bibitem[\protect\citeauthoryear{Evans et al.}{2009}]{Evans09} Evans, P. A., Beardmore, A. P., Page, K. L., et al. 2009, \mnras, 397, 1177
\bibitem[\protect\citeauthoryear{Fan}{2008}]{Fan08} Fan, Y. Z. 2008, \mnras, 389, 1306
\bibitem[\protect\citeauthoryear{Fan}{2009}]{Fan09} Fan, Y. Z. 2009, \mnras, 397, 1539
\bibitem[\protect\citeauthoryear{Fan et al.}{2002}]{Fan02} Fan, Y. Z., Dai, Z. G., Huang, Y. F., \& Lu, T. 2002, ChJAA, 2, 449
\bibitem[\protect\citeauthoryear{Fan \& Wei}{2005}]{Fan05} Fan, Y. Z., \& Wei, D. M. 2005, \mnras, 364, L42
\bibitem[\protect\citeauthoryear{Fan et al.}{2005a}]{Fan05a} Fan, Y. Z., Zhang, B., \& Proga, D. 2005a, \apj, 635, L129
\bibitem[\protect\citeauthoryear{Fan et al.}{2005b}]{Fan05b} Fan, Y. Z., Zhang, B., \& Wei, D. M. 2005b, \apj, 628, L25
\bibitem[\protect\citeauthoryear{Filgas et al.}{2011}]{Filgas11} Filgas, R., Greiner, J., Schady, P., et al. 2011, \aap, 535, A57
\bibitem[\protect\citeauthoryear{Fox et al.}{2003}]{Fox03} Fox, D. W., Price, P. A., Soderberg, A. M., et al. 2003, \apj, 586, L5
\bibitem[\protect\citeauthoryear{Galama et al.}{1998}]{Galama98} Galama, T. J., Vreeswijk, P. M., van Paradijs, J., et al. 1998, \nat, 395, 670
\bibitem[\protect\citeauthoryear{Gao \& Fan}{2006}]{Gao06} Gao, W. H., \& Fan, Y. Z. 2006, ChJAA, 6, 513
\bibitem[\protect\citeauthoryear{Gehrels et al.}{2004}]{Gehrels04} Gehrels, N., Chincarini, G., Giommi, P., et al. 2004, \apj, 611, 1005
\bibitem[\protect\citeauthoryear{Ghirlanda et al.}{2013}]{Ghirlanda13} Ghirlanda, G., Ghisellini, G., Salvaterra, R., et al. 2013, \mnras, 428, 1410
\bibitem[\protect\citeauthoryear{Gorosabel et al.}{2009}]{Gorosabel09} Gorosabel, J., Kubanek, P., Jelinek, M., de Ugarte Postigo, A.,  \& Aceituno, J. 2009, GCN, 9236
\bibitem[\protect\citeauthoryear{Guetta \& Della Valle}{2007}]{Guetta07} Guetta, D., \& Della Valle, M. 2007 \apj, 657, L73
\bibitem[\protect\citeauthoryear{Guidorzi et al.}{2009}]{Guidorzi09} Guidorzi, C., Bersier, D., \& Tanvir, N. 2009, GCN, 9238
\bibitem[\protect\citeauthoryear{G\"otz et al.}{2009}]{Gotz09} G\"otz, D., Laurent, P., Lebrun, F., Daigne, F. \& Bo\v snjak, \v Z. 2009, \apj, 695, L208
\bibitem[\protect\citeauthoryear{Hjorth \& Bloom}{2012}]{Hjorth12} Hjorth, J., \& Bloom, J. 2012, in Gamma-Ray Bursts, ed. C. Kouveliotou, R. A. M. J. Wijers \& S. Woosley (Cambridge Astrophysics Series 51; Cambridge: Cambridge Univ. Press), chap. 9 
\bibitem[\protect\citeauthoryear{Hjorth et al.}{2006}]{Hjorth06} Hjorth, J., Levan, A., Tanvir, N., et al. 2006, The Messenger, 126, 16
\bibitem[\protect\citeauthoryear{Hjorth et al.}{2012}]{Hjorth12} Hjorth, J., Malesani, D., Jakobsson, P., et al. 2012, \apj, 756, 187
\bibitem[\protect\citeauthoryear{Hjorth et al.}{2003}]{Hjorth03} Hjorth, J., Sollerman, J., M\o ller, P., et al. 2003, \nat, 423, 847
\bibitem[\protect\citeauthoryear{Im et al.}{2009a}]{Im09a} Im, M., Park, W., Jeon, Y. et al. 2009a, GCN, 9248 
\bibitem[\protect\citeauthoryear{Im et al.}{2009b}]{Im09b} Im, M., Park, W., Jeon, Y. et al. 2009b, GCN, 9253 
\bibitem[\protect\citeauthoryear{Jin \& Fan}{2007}]{Jin07} Jin, Z. P., \& Fan, Y. Z. 2007, \mnras, 378, 1043
\bibitem[\protect\citeauthoryear{Jin et al.}{2009}]{Jin09} Jin, Z. P., Xu, D., Covino, S., et al. 2009, \mnras, 400, 1829
\bibitem[\protect\citeauthoryear{Li et al.}{2003}]{Li03} Li, W. D., Filippenko, A. V., \& Chornock, R., Jha, S. 2003, \apj, 586, L9
\bibitem[\protect\citeauthoryear{Kann et al.}{2010}]{Kann10} Kann, D.A., Klose, S., Zhang, B., et al. 2010, \apj, 720, 1513
\bibitem[\protect\citeauthoryear{Kewley \& Ellison}{2008}]{Kewley08} Kewley, L. J., \& Ellison, S. L. 2008, \apj, 681, 1183
\bibitem[\protect\citeauthoryear{Kinney et al.}{1996}]{Kinney96} Kinney, A. L., Calzetti, D., Bohlin, R. C., et al. 1996, \apj, 467, 38
\bibitem[\protect\citeauthoryear{Kobayashi}{2000}]{Kobayashi00} Kobayashi, S. 2000, \apj, 545, 807
\bibitem[\protect\citeauthoryear{Malesani et al.}{2004}]{Malesani04} Malesani, D., Tagliaferri, G., Chincarini, G., et al. 2004, \apj, 609, L5
\bibitem[\protect\citeauthoryear{Mao et al.}{2009}]{Mao09} Mao, J., Cha, G., \& Bai, J. 2009, GCN, 9305 
\bibitem[\protect\citeauthoryear{Markwardt et al.}{2008}]{Markwardt08} Markwardt, C. M., Barthelmy, S. D., Baumgartner, W. H., et al. 2008, GCN, 8338
\bibitem[\protect\citeauthoryear{Mazzali et al.}{2006}]{Mazzali06} Mazzali, P. A.; Deng, J., Nomoto, K., et al. 2006, \nat, 442, 1018
\bibitem[\protect\citeauthoryear{Mazzali et al.}{2003}]{Mazzali03} Mazzali, P. A., Deng, J., Tominaga, N., et al. 2003, \apj, 599, L95
\bibitem[\protect\citeauthoryear{Mazzali et al.}{2000}]{Mazzali00} Mazzali, P. A., Iwamoto, K., \& Nomoto, K. 2000, \apj, 545, 407
\bibitem[\protect\citeauthoryear{M{\'e}sz{\'a}ros}{2006}]{Meszaros06} M{\'e}sz{\'a}ros, P. 2006, RPPh, 69, 2259
\bibitem[\protect\citeauthoryear{M{\'e}sz{\'a}ros \& Rees}{1997}]{Meszaros97} M{\'e}sz{\'a}ros, P., \& Rees, M. J. 1997, \apj, 476, 232
\bibitem[\protect\citeauthoryear{M{\'e}sz{\'a}ros et al.}{1998}]{Meszaros98} M{\'e}sz{\'a}ros, P., Rees, M. J., \& Wijers, R. A. M. J. 1998, \apj, 499, 301
\bibitem[\protect\citeauthoryear{Molinari et al.}{2007}]{Molinari07} Molinari, E., Vergani, S. D., Malesani, D., et al. 2007, \aap, 469, L13
\bibitem[\protect\citeauthoryear{Nissinen \& Hentunen }{2009}]{Nissinen09} Nissinen, M., \& Hentunen, V. 2009, GCN, 9246 
\bibitem[\protect\citeauthoryear{Ofek et al.}{2013}]{Ofek13} Ofek, E.O., Sullivan, M., Cenko, S. B., et al. 2013, \nat, 494, 65
\bibitem[\protect\citeauthoryear{Oksanen}{2009}]{Oksanen09} Oksanen, A. 2009, GCN, 9239 
\bibitem[\protect\citeauthoryear{Olivares et al.}{2009}]{Olivares09} Olivares, F., Kupcu Yoldas, A., Greiner, J., \& Yoldas, A. 2009, GCN, 9245  
\bibitem[\protect\citeauthoryear{Page et al.}{2011}]{Page11}Page, K. L., Starling, R. L. C., Fitzpatrick, G., et al. 2011, \mnras, 416, 2078
\bibitem[\protect\citeauthoryear{Panaitescu \& Kumar}{2001}]{Panaitescu01}Panaitescu, A., \& Kumar, P. 2001, \apj, 554, 667
\bibitem[\protect\citeauthoryear{Pastorello et al.}{2007}]{Pastorello07} Pastorello, A., Smartt, S. J., Mattila, S., et al. 2007, \nat, 447, 829
\bibitem[\protect\citeauthoryear{Patat et al.}{2001}]{Patat01} Patat F., Cappellaro, E., Danziger, J., et al. 2001, \apj, 555, 900
\bibitem[\protect\citeauthoryear{Pettini \& Pagel}{2004}]{Pettini04} Pettini, M., \& Pagel, B. E. J. 2004, \mnras, 348, L59
\bibitem[\protect\citeauthoryear{Pian et al.}{2000}]{Pian00}Pian E., Amati, L., Antonelli, L. A., et al. 2000, \apj, 536, 778
\bibitem[\protect\citeauthoryear{Pian et al.}{2006}]{Pian06}Pian E., Mazzali, P. A., Masetti, N., et al. 2006, \nat, 442, 1011
\bibitem[\protect\citeauthoryear{Piran}{1999}]{Pir99}Piran T. 1999, PhR, 314, 575
\bibitem[\protect\citeauthoryear{Racusin et al.}{2008}]{Racusin08}Racusin, J. L., Karpov, S. V., Sokolowski, M., et al. 2008, \nat, 455, 183
\bibitem[\protect\citeauthoryear{Rees \& M{\'e}sz{\'a}ros}{1992}]{Rees92} Rees, M. J., \& M{\'e}sz{\'a}ros, P. 1992, \mnras, 258, 41
\bibitem[\protect\citeauthoryear{Roy et al.}{2009}]{Roy09} Roy, R., Kumar, B., Pandey, S. B., \& Kumar, B. 2009, GCN, 9278
\bibitem[\protect\citeauthoryear{Ryde et al.}{2010}]{Ryde10} Ryde, F., Axelsson, M., Zhang, B. B., et al. 2009, 2010, \apj, 709, L172
\bibitem[\protect\citeauthoryear{Rumyantsev et al.}{2009}]{Rumyantsev09} Rumyantsev, V.,  Antoniuk, K., \& Pozanenko, A. 2009, GCN, 9320
\bibitem[\protect\citeauthoryear{Sakamoto et al.}{2009}]{Sakamoto09} Sakamoto, T., Barthelmy, S. D., Baumgartner, W. H., et al. 2009, GCN, 9231
\bibitem[\protect\citeauthoryear{Sakamoto et al.}{2011}]{Sakamoto11} Sakamoto, T., Barthelmy, S. D., Baumgartner ,W. H., et al. 2011, \apjs, 195, 2
\bibitem[\protect\citeauthoryear{Sari}{1998}]{Sari98} Sari, R. 1998, \apj, 494, L49
\bibitem[\protect\citeauthoryear{Sari \& Piran}{1999}]{Sari99a} Sari, R., \& Piran, T. 1999, \apj, 517, L109
\bibitem[\protect\citeauthoryear{Sari et al.}{1999}]{Sari99} Sari, R., Piran, T., \& Halpern, J. P. 1999, \apj, 524, L43
\bibitem[\protect\citeauthoryear{Savaglio et al.}{2009}]{Savaglio09} Savaglio, S., Glazebrook, K., \& Le Borgne, D. 2009, \apj, 691, 182
\bibitem[\protect\citeauthoryear{Schady et al.}{2012}]{Schady12} Schady, P., Dwelly, T., Page, M. J., et al. 2012, \aap 537, A15
\bibitem[\protect\citeauthoryear{Schlegel et al.}{1998}]{Schlegel98} Schlegel, D. J., Finkbeiner, D. P., \& Davis, M. 1998, \apj, 500, 525
\bibitem[\protect\citeauthoryear{Schulze et al.}{2011}]{Schulze11} Schulze, S., Klose, S., Bj\"ornsson, G., et al. 2011, \aap, 526, A23
\bibitem[\protect\citeauthoryear{Shao \& Dai}{2005}]{Shao05}Shao, L., \& Dai, Z. G., 2005, \apj, 633, 1027
\bibitem[\protect\citeauthoryear{Smith et al.}{2008}]{Smith08}Smith, R. J., Melandri, A., Steele, I. A., et al. 2008 GCN, 8333
\bibitem[\protect\citeauthoryear{Soderberg et al.}{2010}]{Soderberg10} Soderberg, A. M., Chakraborti, S, Pignata, G., et al. 2010, \nat, 463, 513
\bibitem[\protect\citeauthoryear{Sparre et al.}{2011}]{Sparre11} Sparre, M., Sollerman, J., Fynbo, J. P. U., et al. 2011, \apj, 735, L24
\bibitem[\protect\citeauthoryear{Stanek et al.}{2003}]{Stanek03} Stanek, K. Z., Matheson, T., Garnavich, P. M., et al. 2003, \apj, 591, L17
\bibitem[\protect\citeauthoryear{Starling et al.}{2011}]{Starling11} Starling, R. L. C., Wiersema, K., Levan, A. J., et al. 2011, \mnras, 411, 2792
\bibitem[\protect\citeauthoryear{Steele et al.}{2009}]{Steele09}Steele, I. A., Mundell, C. G., Smith, R. J., Kobayashi, S., \& Guidorzi, C. 2009, \nat, 462, 767
\bibitem[\protect\citeauthoryear{Tanvir et al.}{2010}]{Tanvir10} Tanvir, N. R., Rol, E., Levan, A. J., et al. 2010, \apj, 725, 625
\bibitem[\protect\citeauthoryear{Taubenberger et al.}{2006}]{Taubenberger06} Taubenberger, S., Pastorello, A., Mazzali, P. A., et al. 2006, \mnras, 371, 1459 
\bibitem[\protect\citeauthoryear{Troja et al.}{2012}]{Troja12} Troja, E., Sakamoto, T., Guidorzi, C., et al. 2012, \apj, 761, 50
\bibitem[\protect\citeauthoryear{Urata et al.}{2009}]{Urata09} Urata, Y., Zhang, Z. W., Wen, C. Y. et al. 2009, GCN, 9240 
\bibitem[\protect\citeauthoryear{Usov}{1994}]{Usov94} Usov, V. V. 1994, \mnras, 267, 1035
\bibitem[\protect\citeauthoryear{West et al.}{2008}]{West08}West, J. P., Schubel, M., Haislip, J. et al. 2008 GCN, 8352
\bibitem[\protect\citeauthoryear{Wren et al.}{2008}]{Wren08}Wren, J., Vestrand, W. T., Wozniak, P. R., Davis, H., \& Norman, B. 2008 GCN, 8337
\bibitem[\protect\citeauthoryear{Xue et al.}{2009}]{Xue09} Xue, R. R., Fan, Y. Z. \& Wei, D. M. 2009, \aap, 498, 671
\bibitem[\protect\citeauthoryear{Yonetoku et al.}{2011}]{Yonetoku11} Yonetoku, D., Murakami, T., Gunji, S. et al. 2011, \apj, 743, L30
\bibitem[\protect\citeauthoryear{Yuan}{2009}]{Yuan09} Yuan, F. 2009, GCN, 9224
\bibitem[\protect\citeauthoryear{Zhang et al.}{2006}]{Zhang06} Zhang, B., Fan, Y. Z., Dyks, J., et al. 2006, \apj, 642, 354
\bibitem[\protect\citeauthoryear{Zhang et al.}{2003}]{Zhang03} Zhang, B., Kobayashi, S., \& M{\'e}sz{\'a}ros, P. 2003, \apj, 595, 950 
\bibitem[\protect\citeauthoryear{Zhang \& M\'esz\'aros}{2004}]{ZhMe04} Zhang, B. \& M\'esz\'aros P., 2004, IJMPA, 19, 2385
\bibitem[\protect\citeauthoryear{Zhang et al.}{2007}]{Zhang07} Zhang, B., Zhang, B. B., Liang, E. W. et al. 2007, \apj, 655, L25 
\bibitem[\protect\citeauthoryear{Zhang et al.}{2011}]{Zhang11} Zhang, B. B., Zhang, B., Liang, E. W. et al. 2011, \apj, 730, 141


\end{thebibliography}
\end{document}